\documentclass[fleqn,usenatbib]{mnras}

\usepackage[T1]{fontenc}

\DeclareRobustCommand{\VAN}[3]{#2}
\let\VANthebibliography\thebibliography
\def\thebibliography{\DeclareRobustCommand{\VAN}[3]{##3}\VANthebibliography}

\usepackage{graphicx}	% Including figure files
\usepackage{amsmath}	% Advanced maths commands
\usepackage{newtxtext,newtxmath}

\newcommand{\velociraptor}{\textsc{VELOCIraptor}\,}

\title[Splashback with FLAMINGO]{Inferring the dark matter splashback radius from cluster gas and observable profiles in the FLAMINGO simulations.}

\author[I. Towler et al.]{
Imogen Towler,$^{1}$\thanks{E-mail: \href{mailto://imogen.towler@manchester.ac.uk}{imogen.towler@manchester.ac.uk}}
Scott T. Kay,$^{1}$
Joop Schaye,$^{2}$
Roi Kugel,$^{2}$
Matthieu Schaller,$^{3,2}$
Joey Braspenning,$^{2}$
\newauthor
Willem Elbers,$^{4}$
Carlos S. Frenk,$^{4}$
Juliana Kwan,$^{5,6}$
Jaime Salcido,$^{5}$
Marcel P. van Daalen,$^{2}$
\newauthor
Bert Vandenbroucke,$^{2}$
and Edoardo Altamura$^{1}$
\\
$^{1}$Jodrell Bank Centre for Astrophysics, School of Physics and Astronomy, The University of Manchester, Manchester M13 9PL, UK\\
$^{2}$Leiden Observatory, Leiden University, PO Box 9513, 2300 RA Leiden, the Netherlands\\
$^{3}$Lorentz Institute for Theoretical Physics, Leiden University, PO box 9506, 2300 RA Leiden, the Netherlands\\
$^{4}$Institute for Computational Cosmology, Department of Physics, University of Durham, South Road, Durham, DH1 3LE, UK\\
$^{5}$Astrophysics Research Institute, Liverpool John Moores University, Liverpool L3 5RF, UK\\
${^6}$Department of Applied Mathematics and Theoretical Physics, University of Cambridge, Wilberforce Road, Cambridge, CB3 0WA, UK\\
}

\date{Accepted XXX. Received YYY; in original form ZZZ}

\pubyear{2023}

\begin{document}
\label{firstpage}
\pagerange{\pageref{firstpage}--\pageref{lastpage}}
\maketitle

\begin{abstract}
The splashback radius, coinciding with the minimum in the dark matter radial density gradient, is thought to be a universal definition of the edge of a dark matter halo. Observational methods to detect it have traced the dark matter using weak gravitational lensing or galaxy number counts. Recent attempts have also claimed the detection of a similar feature in Sunyaev-Zel'dovich (SZ) observations of the hot intracluster gas. Here, we use the FLAMINGO simulations to investigate whether an extremum gradient in a similar position to the splashback radius is predicted to occur in the cluster gas profiles. We find that the minimum in the gradient of the stacked 3D gas density and pressure profiles, and the maximum in the gradient of the entropy profile, broadly align with the splashback feature though there are significant differences. While the dark matter splashback radius varies with specific mass accretion rate, in agreement with previous work, the radial position of the deepest minimum in the log-slope of the gas density is more sensitive to halo mass. In addition, we show that a similar minimum is also present in projected 2D pseudo-observable profiles: emission measure (X-ray); Compton-$y$ (SZ) and surface mass density (weak lensing). We find that the latter traces the dark matter results reasonably well albeit the minimum occurs at a slightly smaller radius. While results for the gas profiles are largely insensitive to accretion rate and various observable proxies for dynamical state, they do depend on the strength of the feedback processes. % These predictions can be tested with upcoming, multi-wavelength cluster surveys to help further our understanding of the behaviour of the gas and dark matter in cluster outskirts.
\end{abstract}


%%%%%%%%%%%%%%%%%%%%%%%%%%%%%%%%%%%%%%%%%%%%%%%%%%

%%%%%%%%%%%%%%%%% BODY OF PAPER %%%%%%%%%%%%%%%%%%

\section{Introduction}

Galaxy clusters are the result of hierarchical structure formation, forming from the collapse of dark matter overdensities. The gravitational potential provided by these massive halos allows gas, galaxies and stars to reside within and form the galaxy clusters we observe. Defining an edge of these systems is not trivial. Currently, spherical overdensities are used, where the boundary radius is defined as the point within which the average density of a cluster reaches a certain value. For example, $R_{\rm{200m}}$ and $R_{\rm{200c}}$ correspond to the radii at which the average cluster density within reaches 200 times the mean and critical density of the universe respectively and $M_{\rm{200m}}$ and $M_{\rm{200c}}$ are the masses within these radii respectively. It has been proposed to instead define a cluster's boundary following the trajectories of dark matter particles. The splashback radius is a boundary between infalling and collapsed dark matter, defined as the radius of the apocentre of the first orbit of dark matter particles \citep{Diemer2014DEPENDENCERATE} and as such is a model-independent physical definition of the cluster edge. It has been shown using simulations that this radius can be identified using the local logarithmic slope of the dark matter density profiles, where the steepest slope corresponds to the dark matter particles piling up at the apocentre of their first orbit \citep{Diemer2017TheCosmology}. \citet{Diemer2020UniversalHalos} found that defining the mass function of a halo using the splashback radius/mass leads to a more universal mass function, i.e. more independent of redshift and cosmology, than using radii such as $R_{\rm{200m}}$. For $\Lambda$CDM models, they find the universality to be similar to using either the virial mass or spherical overdensities. However, in some alternative cosmologies, the splashback mass functions are significantly more universal. Therefore, the splashback radius is likely to be a more meaningful halo boundary definition.

The splashback radius has been detected observationally using a variety of methods. Often, this is done using a stacked sample of galaxy clusters by using a proxy to measure the mass content of the cluster, for example via weak lensing \citep[e.g.][]{Chang2018TheProfiles, Contigiani2019WeakClusters, Shin2021TheClusters, Fong2022FirstLensing}. In addition, the splashback radius has been detected using measurements of the galaxy number density \citep[e.g.][]{More2016DetectionClusters, Baxter2017TheSDSS, Zurcher2019Clusters, Shin2021TheClusters, Kopylova2022TheRedshifts, Rana2023TheSurvey}. However, past works have found that the method of galaxy cluster selection affects the obtained splashback radius \citep[e.g.][]{Busch2017AssemblyClusters}, with optically selected clusters resulting in smaller splashback radii than expected from simulations \citep{More2016DetectionClusters, Chang2018TheProfiles, Shin2019MeasurementACT}. Currently, there has only been one tentative measurement of the splashback radius of an individual cluster rather than a stacked sample, measured using intracluster light \citep{Gonzalez2021DiscoveryJ1149.5+2223}.

Simulations, both $N$-body and hydrodynamical, have been used to investigate the splashback radius. They allow both the measurement of the trajectories of the individual dark matter particles as well as of the slope of the density profiles to determine the splashback radius, which \citet{Diemer2014DEPENDENCERATE} showed are the same feature. Investigation has led to the well known negative correlation between specific mass accretion rate and the splashback radius of a cluster (normalised using the spherical overdensity radius, $R_{\rm{200m}}$) \citep[e.g.][]{Diemer2014DEPENDENCERATE, Diemer2017TheCosmology, Mansfield2017SplashbackHalos, Deason2021StellarLight, ONeil2021TheSimulations, Diemer2022AResults}. In addition, \citet{Diemer2017TheCosmology} and \citet{ONeil2021TheSimulations} (the latter of which used IllustrisTNG, a full hydro simulation) have both shown further correlations between the splashback radius of a cluster and its mass and redshift. Furthermore, simulations allow the investigation of projection effects on the obtained splashback radius. For example, \citet{Deason2021StellarLight}, who studied both dark matter and stellar density profiles in simulated clusters, found that on average the location of the caustic in projected profiles is approximately ten per cent smaller than the caustic found in 3D density profiles.

%It may be possible to measure the splashback radius using gas observables. For example, \citet{Shi2016LocationsRates}, show that the accretion shock radius in the gas is predicted to align with the splashback radius in the spherical collapse model. However, \citet{ONeil2021TheSimulations} have shown that the minima extracted from log-slope gas density profiles occur at consistently smaller radii than the splashback radii obtained from log-slope dark matter density profiles. In addition, the gas observable would have to be carefully selected because the minima of gas properties such as the temperature, pressure and entropy may not only occur at the splashback radius but also at larger radii due to accretion shocks \citep{Aung2021ShockClusters}. Therefore, SZ observables may be more useful to measure shock accretion radii than the splashback radii \citep[as shown in][]{Baxter2021ShocksSimulations}. However, \citet{Anbajagane2023CosmologicalACT} attempt to measure the locations of shock features observationally in Compton-$y$ profiles and find a minimum in the log-slope profile in the region we expect to find the splashback radius.

It may also be possible to infer the position of the dark matter splashback radius from gas observables. \citet{Lau2015MASSCLUSTERS} and \citet{Aung2021ShockClusters} studied a set of 65 clusters from the \textit{Omega500} non-radiative cosmological simulation \citet{Nelson2014WeighingClusters}, finding the location of the minimum gas density gradient (and maximum entropy gradient) to occur around the same location as the splashback radius. \citet{ONeil2021TheSimulations} studied the Illustris-TNG300 simulation \citep[e.g.][]{Nelson2018FirstBimodality} and showed that the gas density gradient minima consistently occurred at smaller radii than for the dark matter profiles (around 20-30 per cent on cluster scales).  Observationally, \citet{Anbajagane2023CosmologicalACT} found a similar feature when analysing SZ Compton-$y$ (i.e. projected gas pressure) profiles of around $10^{5}$ clusters.

As the gas is collisional, it does not behave in the same way as the dark matter so any physical origin of the association between their density gradient minima is unclear. One possibility is that the gas feature is associated with the accretion shock. \citet{Shi2016LocationsRates} studied self-similar spherical collapse models \citep{Bertschinger1985Self-similarUniverse} and showed that this radius coincides with the splashback radius in clusters with $\gamma=5/3$ and moderate accretion rates. However, \citet{Aung2021ShockClusters}, who locate the shock radius using the minimum entropy gradient, find cluster accretion shocks at around twice the splashback radius \citep[a similar result was found by][based on SZ Compton-$y$ profiles]{Baxter2021ShocksSimulations}. An alternative possibility is that the gas profile shape is the result of the underlying (dark matter-dominated) gravitational potential.

In this work we use the FLAMINGO simulations \citep{Schaye2023TheSurveys, Kugel2023FLAMINGO:Learning} to investigate methods of identifying the splashback radius in galaxy clusters, focusing particularly on identifying a reflection of the splashback feature in baryonic profiles and whether it is possible to find a corresponding splashback feature in potentially observable projected profiles. The FLAMINGO simulation suite contains cosmological boxes run with full hydrodynamics up to a box of side length 2.8 Gpc. This results in hundreds of thousands of simulated galaxy clusters with $M_{\rm{200m}} > 10^{14} \rm{M_{\odot}}$, giving an excellently sized sample to determine the best way to identify the splashback radius and potentially its reflection in cluster gas properties.

In Section \ref{sec:sims}, we summarise the FLAMINGO simulations and the different baryonic models we have analysed in this work. We also define the profiles we have extracted from the simulations, both 3D density profiles and 2D projected observable profiles. In Section \ref{sec:results}, we discuss stacking the cluster profiles. Next, we discuss the minima obtained from the log-slope of various cluster profiles, how these depend on cluster properties and how well the minima in the dark matter and gas densities correspond. Furthermore, in Section \ref{sec:models}, we discuss how the correspondence between the minima in the dark matter and gas densities varies between simulations with different baryonic physics and cosmological models. In Section \ref{sec:projection}, we present our results for projected observables, both gas and dark matter, and investigate which give a good estimate for the splashback radius. Finally in Section \ref{sec:conclusions}, we summarise our results.

\section{FLAMINGO Simulations}
\label{sec:sims}

The FLAMINGO simulations \citep[Full-hydro Large-scale structure simulations with All-sky Mapping for the Interpretation of Next Generation Observations][]{Schaye2023TheSurveys} are a suite of cosmological simulations for cluster physics and cosmology run using the Smoothed Particle Hydrodynamics (SPH) code SWIFT \citep{Schaller2023Swift:Applications}. The suite contains a variety of cosmological simulation boxes with side lengths up to 5.6 Gpc when running dark matter only and 2.8 Gpc with full hydrodynamics. 

FLAMINGO bases its subgrid prescriptions on those developed for the OWLS \citep{Schaye2010} and EAGLE \citep{SchayeTheEnvironments2015} projects. This includes element-by-element radiative cooling and heating rates from \citet{Ploeckinger2020RadiativeFields}; stellar mass loss from stellar winds arising from core-collapse supernovae, type Ia supernovae, massive stars and asymptotic giant branch stars implemented as described in \citet{Wiersma2009ChemicalSimulations} and \citet{SchayeTheEnvironments2015}; stellar feedback as in \citet{DallaVecchia2008SimulatingFeedback} and \citet{Chaikin2023AFormation}, which is implemented by kicking SPH neighbours of young star particles; placing black hole seeds in sufficiently massive regions following \citet{DiMatteo2008DirectGalaxies} and \citet{Booth2009}; and thermally implemented AGN feedback following \citet{Booth2009}.

A variety of models were run in addition to the fiducial model in a 1 Gpc box, this includes 8 alternative astrophysics variations and 4 additional cosmologies. The astrophysical variations were calibrated to different values of the low-redshift galaxy stellar mass function and galaxy cluster gas fractions \citep[See Table 2  of][for details of the variations of the observable data]{Schaye2023TheSurveys}. Varying four of the subgrid parameters allowed these observed quantities to be altered in the resulting simulation by fixed amounts \citep[the variations of these four parameters are given in Table 1 of][]{Schaye2023TheSurveys}. In Section \ref{sec:models}, in addition to the fiducial model, we look at results from the alternative astrophysics models that vary the cluster gas fraction (labelled as fgas+2, -2, -4 and -8$\sigma$). The subgrid parameters of the fiducial model were calibrated directly to observations whereas the different fgas models were calibrated to match observed errorbars of the cluster gas fraction using machine learning optimisation \citep{Kugel2023FLAMINGO:Learning}. In addition, we also look at the two models that alter the AGN feedback mechanism from thermal injection to jet feedback (Jet and Jet\_fgas-4$\sigma$) using the method of \citet{Husko2022Spin-drivenClusters}. These were separately calibrated to observed data or their errorbars. This results in more energy output from AGN feedback being distributed to the outskirts of clusters, albeit non-isotropically.

In addition to the effect of the baryonic model, we also briefly look at the cosmology variations of the simulation (see Appendix \ref{app:cosmo}). The fiducial model uses the cosmological model given by the Dark Energy Survey Y3 \citep{DESCollaboration2022DarkLensing} and the alternative models use results from Planck \citep{PlanckCollaboration2020PlanckResults} or the "lensing cosmology" from \citet{Amon2023ConsistentKiDS-1000} and include varying neutrino masses.

In this work we analyse the results of two simulation box sizes: 1.0 and 2.8 Gpc with a resolution giving a particle gas mass of $m_{\rm{gas}} \approx 10^9\, \rm{M}_{\odot}$, these are labelled as L1\_m9 and L2p8\_m9 respectively. The larger of these was only run with the fiducial model. We use data from the smaller box to compare the different cosmological models and baryonic physics runs as well as to investigate the effects of projection. In both cases, our sample is selected such that all clusters have a mass of $M_{\rm{200m}} > 10^{14}\,  \rm{M_{\odot}}$.

\subsection{Profile definitions}
\label{sec:sims:profiles}

In this work, profiles were extracted from galaxy clusters in the FLAMINGO simulation data set. 3D density profiles for both gas and dark matter ($\rho_{\rm{gas}}$ and $\rho_{\rm{DM}}$ respectively) were obtained by centring a series of spherical shells on the \velociraptor \citep{Elahi2019HuntingVELOCIRAPTOR} determined halo centre within 0.1 - 5 $R_{\rm{200m}}$ with 44 equally spaced logarithmic bins and measuring the total mass of each particle type in that shell. The density profiles were extracted for all clusters with a mass $M_{\rm{200m}} > 10^{14}\,  \rm{M_{\odot}}$ (defined as the mass within $R_{\rm{200m}})$, giving approximately 16,000 clusters for each model in the 1 Gpc box and 380,000 clusters in the 2.8 Gpc box. In addition to the 3D density profiles for gas and dark matter, we also measure the mass-weighted gas temperature,
\begin{equation}
    T = \frac{\sum_{i} m_{i} T_{i}}{\sum_{i} m_{i}},
\end{equation}
where we weight the temperature by the mass ($m$) of the $i^{\rm th}$ particle.
We use the gas density and temperature profiles to estimate the pressure, $P$, and entropy, $K$, profiles as
\begin{equation}
    P = \frac{\rho_{\rm{gas}}}{\mu m_{\rm{p}}} k_{\rm{B}} T,
\end{equation}
\begin{equation}
    K = \frac{k_{\rm{B}} T}{(\rho_{\rm{gas}}/\mu m_{\rm{p}})^{2/3}},
\end{equation}
where we assume homogeneous clusters with primordial abundance giving $\mu=0.59$ as the mean molecular weight.

In addition to the 3D profiles, we extract potentially observable 2D profiles to investigate how well we can obtain the splashback radius from different observables. These include the total mass surface density profile (relevant to weak lensing), hot gas emission measure (a proxy for the soft X-ray band) and integrated Compton-$y$ profiles (SZ). For each of these, a series of cylindrical shells were placed around the cluster centre of potential with a total depth of 10 $R_{\rm{200m}}$, this depth was found to be sufficiently deep to ensure the profiles were converged. Each cylindrical bin is split into $N_{\rm{seg}}=50$ angular segments and a median value for each radial bin is calculated \citep[more detail of this process is in][]{Towler2022GasSimulations} to reduce the noise in the profile due to substructures \citep{Mansfield2017SplashbackHalos, Deason2021StellarLight}. While this azimuthal median method has been used observationally in X-rays, the splashback radius is located in the very outskirts of clusters, where the signal to noise is limited. Therefore, it will only be possible to measure observable profiles using this technique in future surveys.
We measure the surface density, $\Sigma$, by calculating the total mass density in each segment and taking a median over all segments (represented by $\langle\rangle_{\rm{seg}}$) in each radial bin,
\begin{equation}
    \Sigma = \left\langle \frac{\sum_{i} m_{i}}{V_{\rm{seg}}} \right\rangle_{\rm{seg}},
\end{equation}
where $V_{\rm{seg}}$ is the volume of the angular segment. The emission measure is computed as
\begin{equation}
    EM(R) = \left\langle \frac{X_{\rm{H}}}{\mu_{\rm{e}}m_{\rm{p}}^2 A_{\rm{seg}}} \sum_{i=1}\rho_{i} m_{i}\right\rangle_{\rm{seg}},
\end{equation}
where $X_{\rm{H}}=0.76$ represents the hydrogen mass fraction, $\mu_{\rm{e}}=1.14$ the mean molecular weight per free electron and, $A_{\rm{seg}}$ the cross-sectional area of the angular segment. Finally, the integrated Compton-$y$ is measured following
\begin{equation}
    y = \left\langle \frac{k_{\rm{B}} \sigma_{\rm{T}}}{m_{e} c^2 \mu_{e} m_{\rm{H}} A_{\rm{seg}}} \sum^{N_{\rm{seg}}}_{i=1} T_{i} m_{i} \right\rangle_{\rm{seg}}.
\end{equation}
We measure these profiles three times for each cluster, one for each perpendicular projection along each simulation axis and each of these are treated as an independent cluster profile.

\subsection{Stacking}
\label{sec:stacking:criteria}

The splashback radius is large enough that observers normally need to stack profiles to be able to identify the splashback feature in the outskirts of clusters. In addition, clusters contain intrinsic scatter in their density profile due to the presence of substructure and so we can improve the noise levels of the extracted density profiles by stacking them and obtain an average profile for similar clusters. We split the cluster sample over bins of equal spacing in the chosen quantity for stacking. Once a set of profiles have been stacked, we smooth the profiles using the fourth order Savitzky-Golay smoothing algorithm \citep{Savitzky1964SmoothingProcedures} with a window size of the 19 nearest bins to remove any remaining noise, but this effect is minimal due to the large sample used. From these profiles, we take the radial gradient of the log profiles and then extract the radius of the gradient minimum in the profile to get the splashback radius or the location of the gas minima. In this section, we discuss the criteria we use to stack, using criteria which are all indicators of cluster dynamical state, and the effect of the cluster selection and stacking on the profiles.

\subsubsection{Theoretical criteria}

It has been shown that the splashback radius is strongly correlated with the accretion rate of the cluster \citep{Diemer2014DEPENDENCERATE}. The gravitational potential in high accretion rate clusters deepens faster, decreasing the splashback radius because it reflects earlier infall. The correlation between these means that stacking in bins of accretion rate leads to profiles with similar splashback radii and so the stacked splashback feature is relatively narrow and will not be broadened due to stacking.

Following \citet{Diemer2017TheCosmology}, we define the specific accretion rate to be 
\begin{equation}
    \Gamma (t) = \frac{ \log \left[M_{\rm{200m}} (t) \right] - \log \left[ M_{\rm{200m}} (t-t_{\rm{dyn}}) \right]} {\log \left [a(t) \right] - \log \left[ a \left(t-t_{\rm{dyn}} \right) \right]},
    \label{eq:accretion}
\end{equation}
where $t_{\rm{dyn}}$ is the dynamical time and corresponds to approximately $a(t-t_{\rm{dyn}})=\frac{2}{3}$ ($z=0.5$) for halos at $a(t)=1$ ($z=0$) \citep{Diemer2017TheAlgorithm, Deason2021StellarLight}.

The ratio of the kinetic and thermal energy ($E_{\rm{kin}}$ and $E_{\rm{therm}}$ respectively) within a cluster can also be used to measure the dynamical state of simulated clusters \citep[e.g.][]{Barnes2017a}. This is measured using only one snapshot of the simulation and so is a more instantaneous measure of the dynamical state of the cluster than the accretion rate, which is measured over a dynamical time. In addition, as it is measured using the energy of the gas, it can capture any dynamical differences that arise in the gas that might not exist in the dark matter, e.g. due to feedback processes. It is calculated via
\begin{equation}
   X_{\rm{E}} = \frac{E_{\rm{kin}}}{E_{\rm{therm}}} = \mu m_{\rm{H}} \frac{\sum_i m_{i} v_{i}^2}{3 k_{B} \sum_{i} m_{i} T_{i}},
   \label{eq:energy}
\end{equation}
summing over the mass, cluster rest frame speed ($v_{i}$) and temperature of the gas particles $i$ within $R_{\rm{200m}}$. Similar to the accretion rate and the true mass of a cluster, this is a purely theoretical quantity and so cannot be derived from observational data.

\begin{figure*}
    \centering
    \includegraphics{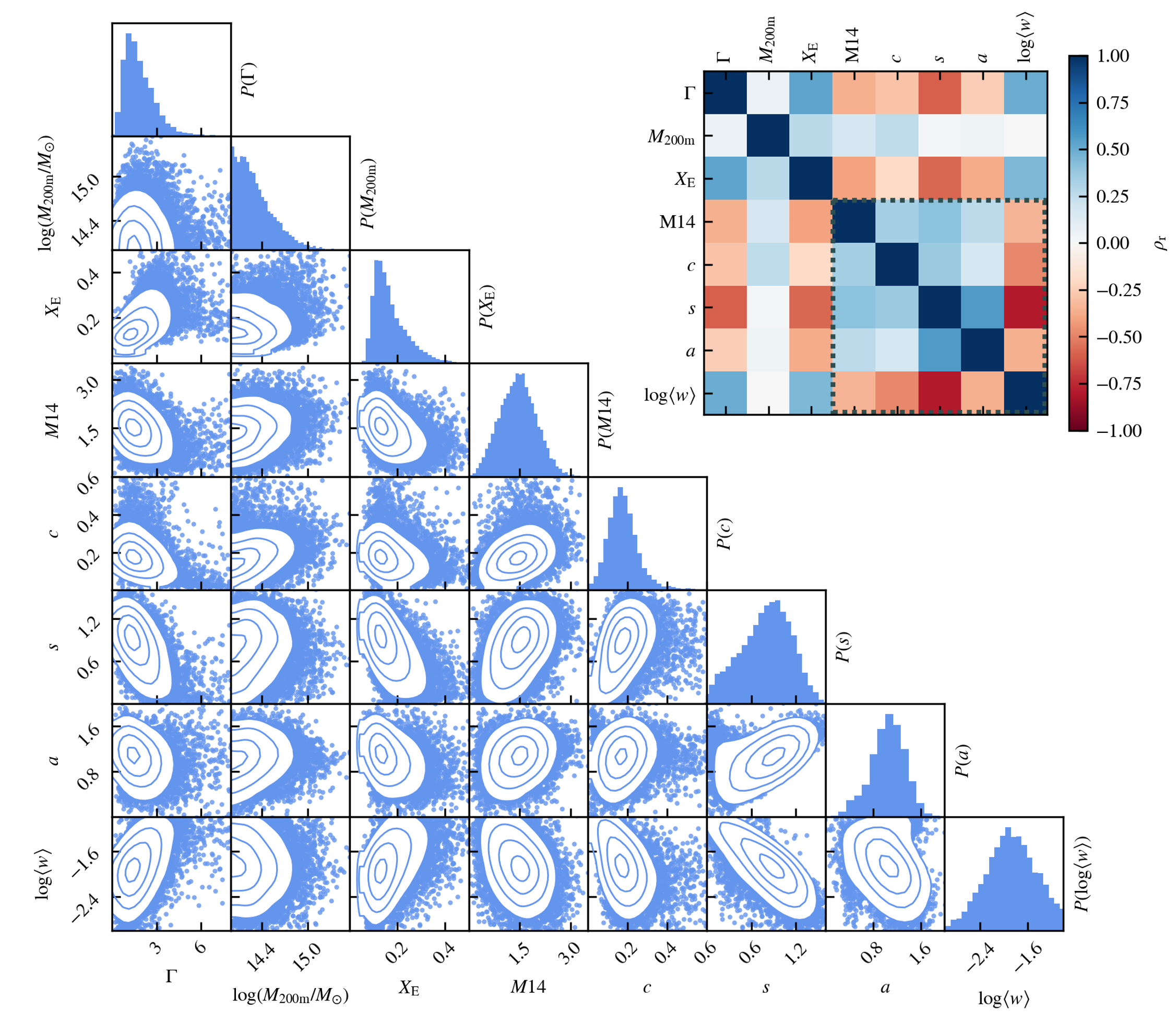}
    \caption{Corner plot comparing the total cluster mass, specific accretion rate (Equation \ref{eq:accretion}), gas energy ratio (Equation \ref{eq:energy}), magnitude gap, concentration (Equation \ref{eq:concentration}), symmetry (Equation \ref{eq:symmetry}), alignment (Equation \ref{eq:alignment}) and centroid shift (Equation \ref{eq:centroid}) for the L1\_m9 cluster sample. The diagonal gives the distribution for each quantity. The upper right corner shows a correlation matrix with the Pearson coefficients for each of the different stacking criteria. The dashed box shows the area containing the correlations between the different 2D criteria.}
    \label{fig:parameter_corner}
\end{figure*}

\subsubsection{Observable criteria}
\label{sec:stacking:obs}

Due to the well known correlation between the splashback radius and the accretion rate \citep[e.g.][]{Diemer2014DEPENDENCERATE, Wetzel2015TheGalaxies, Mansfield2017SplashbackHalos, ONeil2021TheSimulations}, we would ideally stack cluster profiles in bins of accretion rate. However, it is not directly observable, so instead we use methods of measuring the current dynamical state of clusters, as those that have recently accreted large amounts of mass, e.g. through mergers, are much more likely to be disturbed. We investigate a series of gas morphology criteria to investigate their correlations with the accretion rate and splashback radius. These were measured using emission measure maps of the clusters and therefore trace the gas distribution within a cluster. We use the following:
\begin{itemize}
    \item The concentration parameter,
    \begin{equation}
        c = \frac{EM(<0.15 R_{500\rm{c}})}{EM(<R_{500\rm{c}})},
        \label{eq:concentration}
    \end{equation}
    which compares the total emission measure within two apertures of $0.15 R_{\rm{500c}}$ and $R_{500\rm{c}}$ to find clusters with a brighter, cooler core, which tend to be more relaxed \citep{Peterson2006} and have a larger $c$.

    \item The symmetry statistic \citep{Mantz2015},
    \begin{equation}
        s = - \log_{10} \left( \frac{1}{N_{\rm{el}}} \sum_{j=1}^{N_{\rm{el}}} \frac{\delta_{j,c}}{\langle b_{\rm{el}} \rangle}_j \right),
        \label{eq:symmetry}
    \end{equation}
    where a series of $N_{\rm{el}}=5$ ellipses have been fit to an emission measure map (a proxy for surface brightness in this case) of a cluster at different brightness levels varying between $0.1 - 1.0 R_{\rm{500c}}$. The distances between the centres of the ellipses and the cluster centre, $\delta_{j,c}$, are compared with the average of the minor and major axes of the $j$th ellipse, $\langle b_{\rm{el}} \rangle_{j}$.  This measures the symmetry of a cluster around its global centre, in this case the point of minimum potential. A higher value of $s$ shows that a cluster is more symmetrical and therefore more relaxed.

    \item The alignment statistic,
    \begin{equation}
        a = - \log_{10} \left( \frac{1}{N_{\rm{el}} - 1} \sum_{j=1}^{N_{\rm{el}}-1} \frac{\delta_{j,j+1}}{\langle b_{\rm{el}} \rangle_{j,j+1}} \right),
        \label{eq:alignment}
    \end{equation}
    which is measured using the same fitted ellipses as the symmetry statistic. However, this parameter aims to measure how the amount of substructure shifts at different radii. Therefore, it instead compares the distances between the centres of adjacent ellipses, $\delta_{j,j+1}$, to the average of the ellipse axes of the same adjacent ellipses, $\langle b_{\rm{el}} \rangle$. Similarly to the symmetry statistic, a larger value of $a$ shows a cluster is more relaxed.

    \item The centroid shift, \citep{Maughan2012}
    \begin{equation}
        \langle w \rangle = \frac{1}{R_{500\rm{c}}} \sqrt{\frac{\sum_{i}^{M}\left( \Delta_{i} - \langle \Delta \rangle \right)^2}{M-1}},
        \label{eq:centroid}
    \end{equation}
    measures the distance between the centroid of the surface brightness and the global centre of the cluster ($\Delta$), averaged over $M=8$ increasingly smaller apertures within $0.15 - 1.0 R_{500c}$. Smaller values of $\langle w \rangle$ correspond to a smaller shift and the clusters are therefore more relaxed.
\end{itemize}

In addition to these morphology criteria, we investigate whether the magnitude gap, the difference in magnitude between the brightest cluster galaxy (BCG) and the $n^{\rm th}$ brightest galaxy, can be used as a proxy for the accretion rate. Over time, satellite galaxies get tidally disrupted and stripped of matter, and it has been shown that larger satellites are affected more by dynamical friction. Therefore, as a halo ages, the brightness gap between the BCG and the brightest satellites grows. \citet{Shin2022WhatProperties} proposed that the magnitude gap should negatively correlate with accretion rate as a large magnitude gap is an indicator of an old halo and hence low accretion rate. Following \citet{Farahi2020AgingHaloes}, we measure the magnitude gap between the BCG and fourth brightest galaxy and denote this as $M14$. We use the galaxy $r$-band luminosities measured within a 50 pkpc 3D aperture around the galaxy centre provided by the Spherical Overdensity and Aperture Processor (SOAP\footnote{SOAP is a tool developed as part of the FLAMINGO project. The code is available at \href{https://github.com/SWIFTSIM/SOAP}{https://github.com/SWIFTSIM/SOAP}}) catalogue to determine the galaxy luminosities and hence measure the magnitude gap of each cluster.

We compare both the theoretical and observational criteria taken from L1\_m9 in Fig. \ref{fig:parameter_corner}, the relationships between quantities are shown in the off diagonals and the distribution of each quantity on the diagonal. In addition, in the top-right, we show a correlation matrix showing the Pearson correlation coefficients between the different criteria. The morphology criteria were calculated three times for each cluster, one for each perpendicular axis from emission measure maps. In Fig. \ref{fig:parameter_corner}, only one direction is chosen to keep the sample sizes the same between 2D and 3D criteria. We find that there is a weak correlation between the mass of a cluster and almost all other quantities. Therefore, when splitting clusters into bins of the other quantities, each bin will be dominated by low-mass clusters. Conversely, we find particularly strong correlations between the accretion rate and the energy ratio, symmetry statistic and centroid shift. In cases with higher mass accretion, we expect the gas within a cluster to be more disturbed and therefore the energy ratio increases with the accretion rate. In addition, both the symmetry statistic and centroid shift measure how visibly dynamically disturbed the cluster is and so we expect these to correlate well with the accretion rate.

\section{Results from 3D profiles}
\label{sec:results}

In this section, we investigate the effects of stacking the 3D profiles obtained from the $z=0$ output of FLAMINGO's fiducial hydro run, the 2.8 Gpc box, L2p8\_m9. We investigate both the stacked dark matter and gas profiles as well as the radius and depth of the minima of the slope found in each profile. 

\subsection{Effects of stacking DM profiles}
\label{sec:stacking_profiles}

\begin{figure*}
    \centering
    \includegraphics{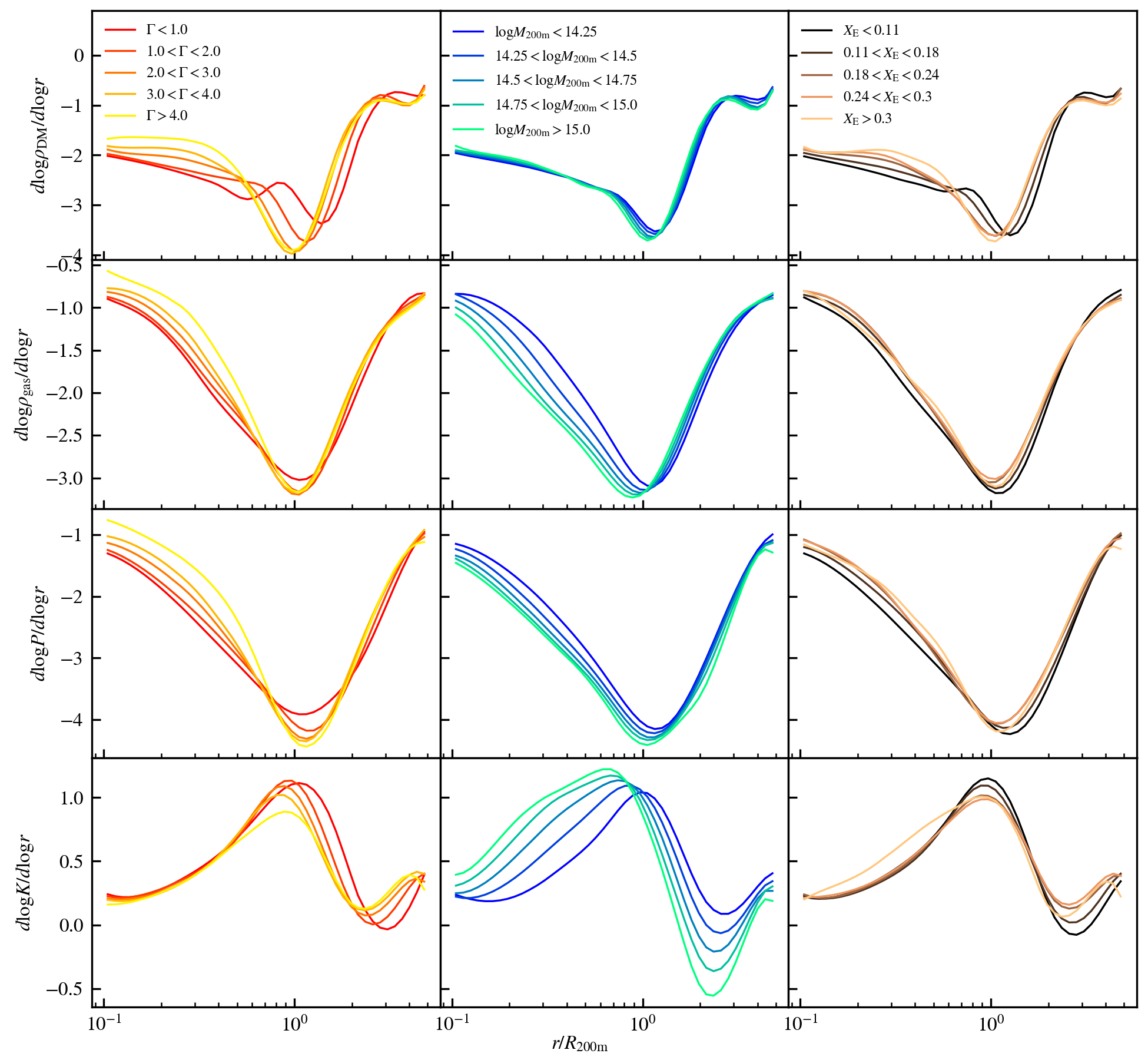}
    \caption{Comparison of the effect of using different stacking criteria on the stacked (from top to bottom) dark matter density, gas density, pressure and entropy gradient profiles.
    Left column: accretion rate, (Equation \ref{eq:accretion}), middle: mass, and right: gas kinetic-thermal energy ratio (Equation \ref{eq:energy}).}
    \label{fig:profile_comparison}. 
\end{figure*}

Fig. \ref{fig:profile_comparison} shows the stacked dark matter density (top row), gas density (second row), pressure (third row) and entropy (bottom row) gradient profiles in bins of accretion rate (left column), mass (middle column) and energy ratio (right column). In agreement with \citet{Diemer2014DEPENDENCERATE, Diemer2017TheCosmology, ONeil2021TheSimulations}, we find that the minimum (most negative) local gradient in the dark matter profile corresponding to the splashback radius depends on the accretion rate of the clusters (see top left panel). This is to be expected as a larger recent accretion rate results in a steeper potential which leads to a smaller splashback radius. We also find that the clusters with the lowest accretion rate have an additional feature in the dark matter profiles at a smaller radius than the splashback radius. \citet{Deason2021StellarLight} suggest this is a ``second caustic'' feature \citep[first discussed in][]{Adhikari2014SplashbackHalos}, corresponding to a build-up of dark matter particles at the apocentre of their second orbit. Clusters with low accretion rates tend to be older and more relaxed, and so the particles will have had enough time to enter their second orbit.

\begin{figure}
    \centering
    \includegraphics[trim = {0 0 0.4cm 1.5cm}, clip]{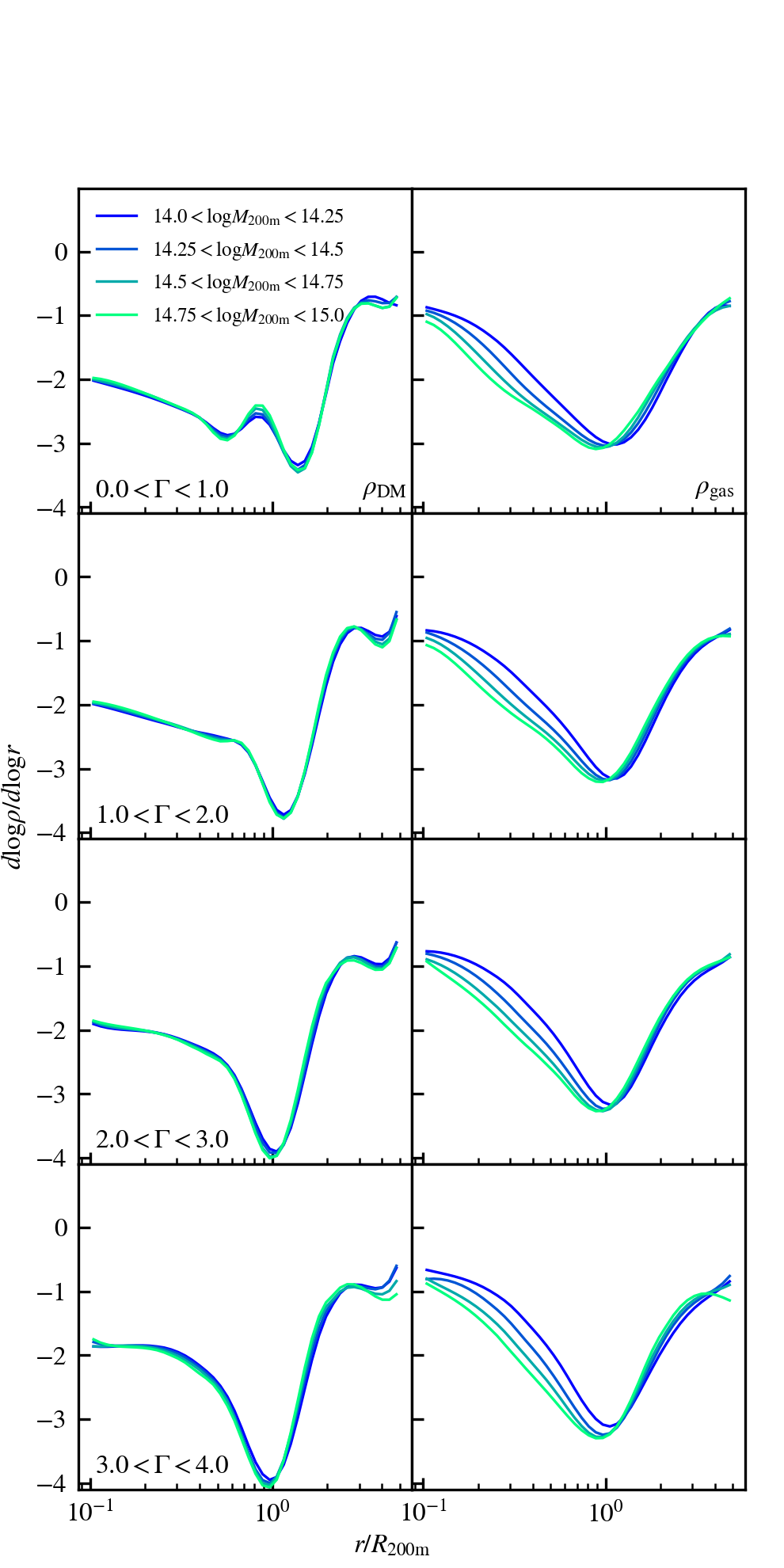}
    \caption{Stacked dark matter (left) and gas (right) density gradient profiles in bins of both accretion rate (Equation \ref{eq:accretion}; different rows) and mass (different colours).}
    \label{fig:grid_mass_acc}
\end{figure}

We also find a weak mass dependence for the splashback radius (see middle column of top row) over the range $M_{\rm{200m}} > 10^{14}\,  \rm{M_{\odot}}$ (where the maximum mass is set by the most massive cluster in our sample, $M_{\rm{200m}} = 10^{15.58}\,  \rm{M_{\odot}}$). This is in agreement with what has been found by \citet{Diemer2017TheCosmology, ONeil2021TheSimulations}. However, the mass dependence of the splashback radius may be due to a correlation between the mass and accretion rate \citep{Diemer2014DEPENDENCERATE} as we expect less massive clusters to be more relaxed. However, from Fig. \ref{fig:parameter_corner}, we expect this correlation to be weak. We investigate this directly by splitting our cluster sample into both mass bins and bins of accretion rate within that. Fig. \ref{fig:grid_mass_acc} shows the resulting dark matter and gas density gradient profiles when stacked in this manner. We find that the dark matter profiles are nearly independent of the cluster mass for a fixed range of accretion rate values. Therefore, it is likely that the small mass dependence in the dark matter profiles of Fig. \ref{fig:profile_comparison} originates entirely from the accretion rate dependence. 

In Fig. \ref{fig:profile_comparison}, we also investigate whether there is a correlation between the splashback radius and the gas kinetic-thermal energy ratio within clusters. In general, the density gradient of the more relaxed clusters, which have a lower fraction of kinetic energy ($X_{\rm{E}}$), match that of the clusters with smaller accretion rates and have a larger splashback radius. In addition, the strong correlation between the mass accretion rate and energy ratio is clear from the fact that the second caustic feature in the least accreting clusters is also visible in the clusters with the lowest energy ratio.

\begin{figure*}
    \centering
    \includegraphics{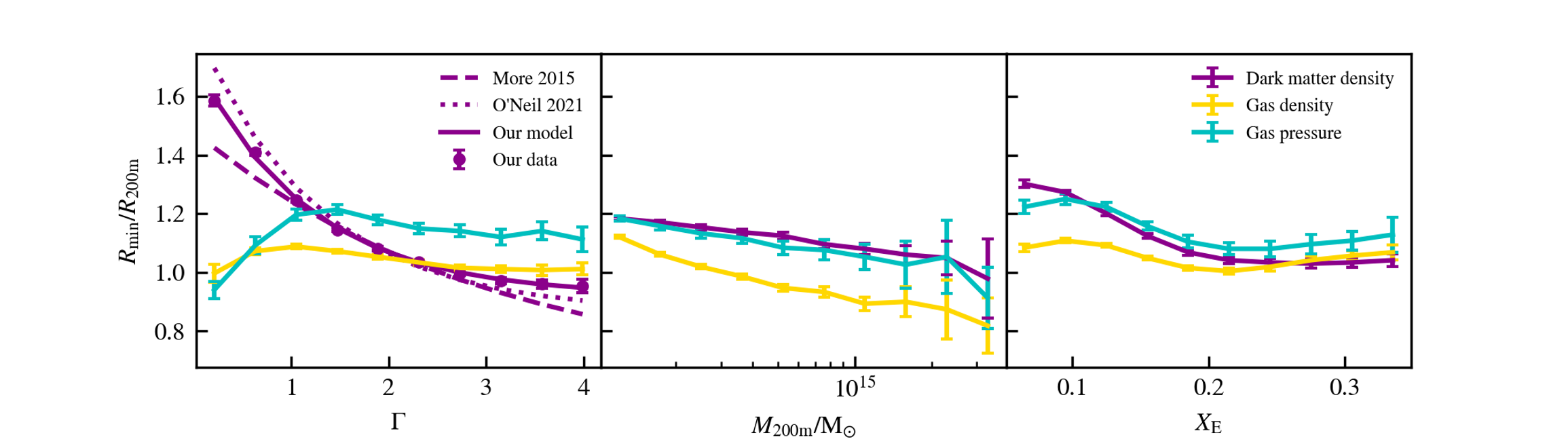}
    \caption{The dependence of the splashback radius on the three properties used to bin the halos, accretion rate (Equation \ref{eq:accretion}, left), mass (centre) and ratio of kinetic and thermal energy of the gas (Equation \ref{eq:energy}, right). The error bars show the three-sigma error found from bootstrapping the cluster sample in each bin used for stacking the profiles.}
    \label{fig:params}
\end{figure*}

Fig. \ref{fig:params} explicitly shows the parameter dependence of the minimum gradient radius (top row) and depth (bottom row) for the dark matter density, gas density and gas pressure profiles. The left panels shows how the radii and depths of the minima depend on the accretion rate. In the upper-left panel, we compare our FLAMINGO results with the \citet{More2015TheMass} model for dark matter density profiles,
\begin{equation}
    R_{\rm{SP}}/R_{\rm{200m}} = A \left[1 + B\Omega_{\rm{m}}(z)\right] \left(1 + C e^{-\Gamma/D}\right),
    \label{eq:accretion_model}
\end{equation}
with the values for the free parameters, $A, B, C$ and $D$ found in \citet{More2015TheMass}, \citet{ONeil2021TheSimulations} and fitted in this work; see Table \ref{table:fits} for the values of the fitted parameters. (However, one should note that \citet{More2015TheMass} uses a slightly different accretion rate definition.) We find that this model fits our relation reasonably well for dark matter. Overall, we find that our results agree with previous results \citep{Diemer2014DEPENDENCERATE, Deason2021StellarLight, ONeil2021TheSimulations}. %We find that the minima in the gas density profiles also have an accretion rate dependence, albeit weaker than for the dark matter. 
In addition, Fig. \ref{fig:params} shows the dependence on mass (central panels) and energy ratio (rightmost panels). %Note, the gas minima have a stronger mass dependence than the dark matter, likely due to impact of feedback also being mass dependent. This is also evident by the minima being shallower in lower mass clusters. 
Typically, the mass dependence of the radius of the splashback feature in the dark matter is attributed to larger halos having a higher accretion rate \citep[e.g.][]{Diemer2014DEPENDENCERATE}, which we also find to be true (see Fig. \ref{fig:grid_mass_acc}).
\begin{table}
    \centering
    \caption{Fitted parameter values of Equation \ref{eq:accretion_model} from \citet{More2015TheMass}, \citet{ONeil2021TheSimulations} and this work.}
    \begin{tabular}{ccccc}
        \hline
        Model & A & B & C & D\\
        \hline
        \citet{More2015TheMass} & 0.54 & 0.53 & 1.36 & 3.04\\
         \citet{ONeil2021TheSimulations} & 0.8 & 0.26 & 1.14 & 1.25\\
         This work & 0.88 & 0.16 & 0.87 & 1.18\\
         \hline
    \end{tabular}
    \label{table:fits}
\end{table}

%\begin{figure*}
%    \centering
%    \includegraphics[trim = {0 0 0 0.4cm}, clip]{Figures/parameter_dependence_gamma.png}
%    \caption{The dependence of the radius (top) and depth (bottom) of the splashback feature on the three properties used to bin the halos: accretion rate (Equation \ref{eq:accretion}, left), mass (centre) and ratio of kinetic and thermal energy of the gas (Equation \ref{eq:energy}, right). The error bars show the three-sigma error found from bootstrapping the cluster sample in each bin used for stacking the profiles.}
%    \label{fig:params}
%\end{figure*}

\subsection{Effects on stacking gas property profiles}

Beyond the accretion shock radius, the gas traces the dark matter component. \citet{Aung2021ShockClusters} found the accretion shock to be 20-100 per cent larger than the splashback radius, when the former is defined as minimum in the log-slope of the entropy. \citet{Farahi2022CorrelationsHalos} also found that the dark matter and gas densities are tightly coupled beyond $R_{\rm{200c}}$ and the correlation between the two quantities weakens within 0.3 $R_{\rm{200c}}$ of the centre of the cluster. Therefore, there is likely a coupling between the gas and dark matter density at scales close to the splashback radius, even if it is not particularly strong. 

While minima in the log-slope of the dark matter density profile are dependent on the orbital dynamics within a halo, the cluster gas is strongly affected by shocks. \citet{Shi2016} finds that for an adiabatic index of $\gamma \approx 5/3$, the self-similar collapse model predicts that the splashback radius and the accretion shock radius align. However, other works investigating the accretion shock radius have found it to be much larger than the range in which we expect to find the splashback radius. As mentioned before, \citet{Aung2021ShockClusters} find that the shock radius is 1.89 times larger than the splashback radius. In addition, \citet{Anbajagane2022ShocksMap} measured the location of minima in stacked observed Compton-$y$ (projected thermal electron pressure) profiles. They find two minima, one at a large radius of 4.58 $R_{\rm{200m}}$, which they find is consistent with accretion shocks seen in other works, and one at 1.08 $R_{\rm{200m}}$, consistent with what \citet{Anbajagane2023CosmologicalACT} found in the SPT and ACT data. They attribute this latter depression to arise from the thermal non-equilibrium between electrons and ions in the intracluster medium. However, they also find that this feature does not appear when creating a comparable stacked sample from The Three Hundred simulations. When including only the most relaxed cluster sample from The Three Hundred simulations, they do find a reproduction of the same minimum even though the simulations do not model non-equilibrium effects between electrons and ions, implying that, in simulations, the existence of this minimum in the extracted Compton-$y$ profile depends on the dynamical state of the cluster sample.
Both shocks and the splashback lead to a drop in the gas density profile \citep{ONeil2021TheSimulations}, but the shocks lead to a wider and shallower minimum when profiles from multiple clusters are stacked, which we see in the right column of Fig. \ref{fig:profile_comparison}, showing the stacked gas density profiles. The radius of the minimum of the gas density slope often matches that of the dark matter, but this feature could be a reflection of the splashback in the gravitational potential or a result of shocks.

In Fig. \ref{fig:grid_mass_acc}, where we split the clusters into bins of both cluster mass and accretion rate, we see that the location of the gas minimum is dependent on both properties (particularly mass), whereas the splashback minimum is solely dependent on the accretion rate. This highlights the effect that the feedback and other baryonic processes are having on the gas density profiles (see also Fig. \ref{fig:params}).

In addition to the gas density, we investigate the logarithmic slope profiles of the gas pressure (left) and entropy (right) in bottom two rows of Fig. \ref{fig:profile_comparison}. We find that the pressure profiles also have a minimum gradient at approximately $R_{\rm{200m}}$, but not a minimum corresponding to the accretion shock radius in the expected range, 2-3 $R_{\rm{200m}}$. The location of the pressure minimum corresponds well with the minima at smaller radii found by \citet{Anbajagane2022ShocksMap} and \citet{Anbajagane2023CosmologicalACT} in observed Compton-$y$ profiles. However, they hypothesise that their pressure deficit arises from a thermal non-equilibrium between ions and electrons, which our simulations do not model. We find minima in the log-slopes of the entropy profile at a radius that corresponds to the expected location of a shock feature, and maxima at approximately the splashback radius. The shapes of the entropy profiles match that  of \citet{Aung2021ShockClusters}, but we find that our minima are much shallower and very strongly dependent on the mass of the cluster. However, we found that the way in which the entropy profiles are constructed affects the results in the outskirts. We combine our temperature and density profiles to obtain our entropy whereas, \citet{Aung2021ShockClusters} calculates the volume-weighted entropy.

Fig. \ref{fig:params} also shows the radii of the minima in the 3D pressure gradient profiles. The parameter dependence of the radius of the minimum of the pressure gradient profiles closely resembles that of the dark matter for the cluster mass and energy ratio.However, when looking at the radii of the minima when stacking according to the cluster mass accretion rate, we find almost no correspondence between the minima in the dark matter density gradient and gas pressure gradient. For example, in the most relaxed clusters ($\Gamma < 1)$ the splashback radius is around 50 per cent larger than the gas pressure (and density) gradient minimum radius.

In summary, we find that the minimum of the log-slope in the gas density and pressure profiles often aligns with that of the dark matter. However, the radii of the minima in the gas properties are particularly dependent on the mass of the cluster, showing that there are other baryonic processes affecting these profiles. Furthermore, we have identified minima in the log-slope of the entropy profiles, corresponding to potential shock features which do not appear in the gas density or pressure profiles.

\subsection{Alternative simulation models}
\label{sec:models}

To investigate the effect of the baryonic physics model on the splashback radius, we extract gas and dark matter density profiles from clusters from simulation runs with alternative physics, detailed in Section \ref{sec:sims}. This includes eight astrophysics variations, which are calibrated to vary the resulting galaxy stellar mass function and galaxy cluster gas fractions at low redshift. Furthermore, there are four alternative cosmology runs including varying neutrino masses\citep[see][for further details]{Schaye2023TheSurveys}, results from these runs are presented in Appendix \ref{app:cosmo}.
 
\subsubsection{Dark-matter-only simulations}

\begin{figure}
    \centering
    \includegraphics{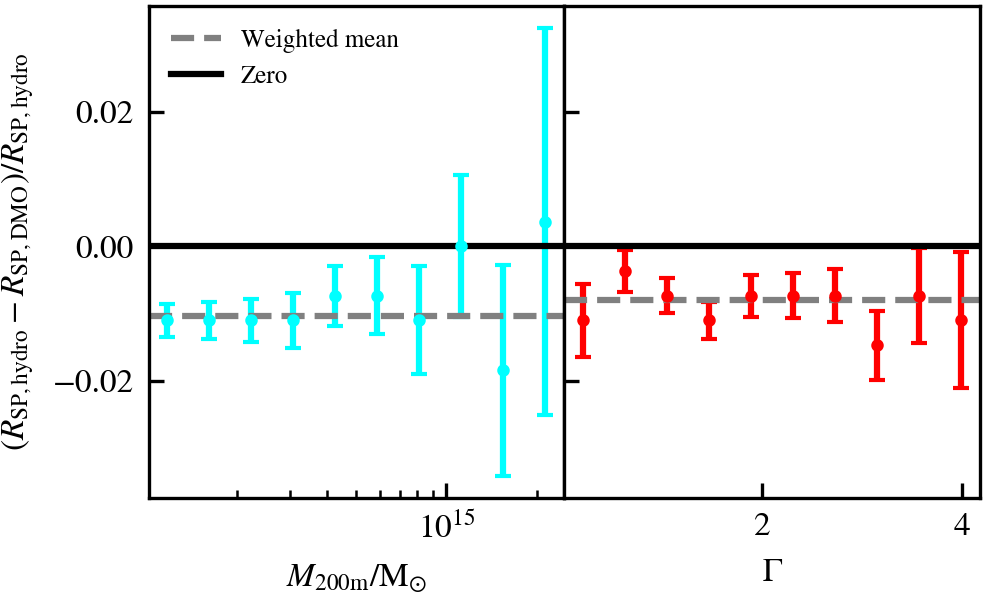}
    \caption{Comparison of the splashback radius extracted from the L1\_m9 and L1\_m9\_DMO simulations when stacking the dark matter density profiles into bins of mass (left) and accretion rate (right). The error bars show the uncertainty by propagating the bootstrap error of the splashback radius from both the DMO and hydro simulations.}
    \label{fig:dmo_scatter}
\end{figure}

Before looking at the different baryonic physics models, we first investigate how the inclusion of baryons affects the dark matter density profiles and recovered splashback radii. We compare the dark matter only run (hereafter DMO), L1\_m9\_DMO, with the hydro L1\_m9 run. The sample chosen from each simulation followed the previous mass cut with all halos with $M_{\rm{200m}} > 10^{14}\, \rm{M_{\odot}}$ included. However, halos tend to be slightly more massive in DMO simulations due to feedback blowing out baryonic matter in the hydrodynamical simulations. Consequently, there are approximately 600 more halos in the DMO sample than the hydro sample, sufficiently small to not have a large effect on the resulting profiles. We highlight the differences in the splashback radii obtained from each set of profiles in Fig. \ref{fig:dmo_scatter}, when stacked according to the mass (left) and accretion rate (right). We find that the differences between the dark matter density profiles for the DMO and hydro runs are minimal. On average, the splashback radius taken from the DMO data is 1 per cent larger than for the same cluster bin in the hydro run. Therefore, the addition of baryons to the simulation has a minimal effect on the presence of the splashback feature in the dark matter, in agreement with \citet{ONeil2021TheSimulations}. The depth of the splashback feature is similarly unaffected, in the DMO clusters the gradient is on average 1 per cent lower than the clusters from the hydro simulation. 

\begin{figure*}
    \centering
    \includegraphics{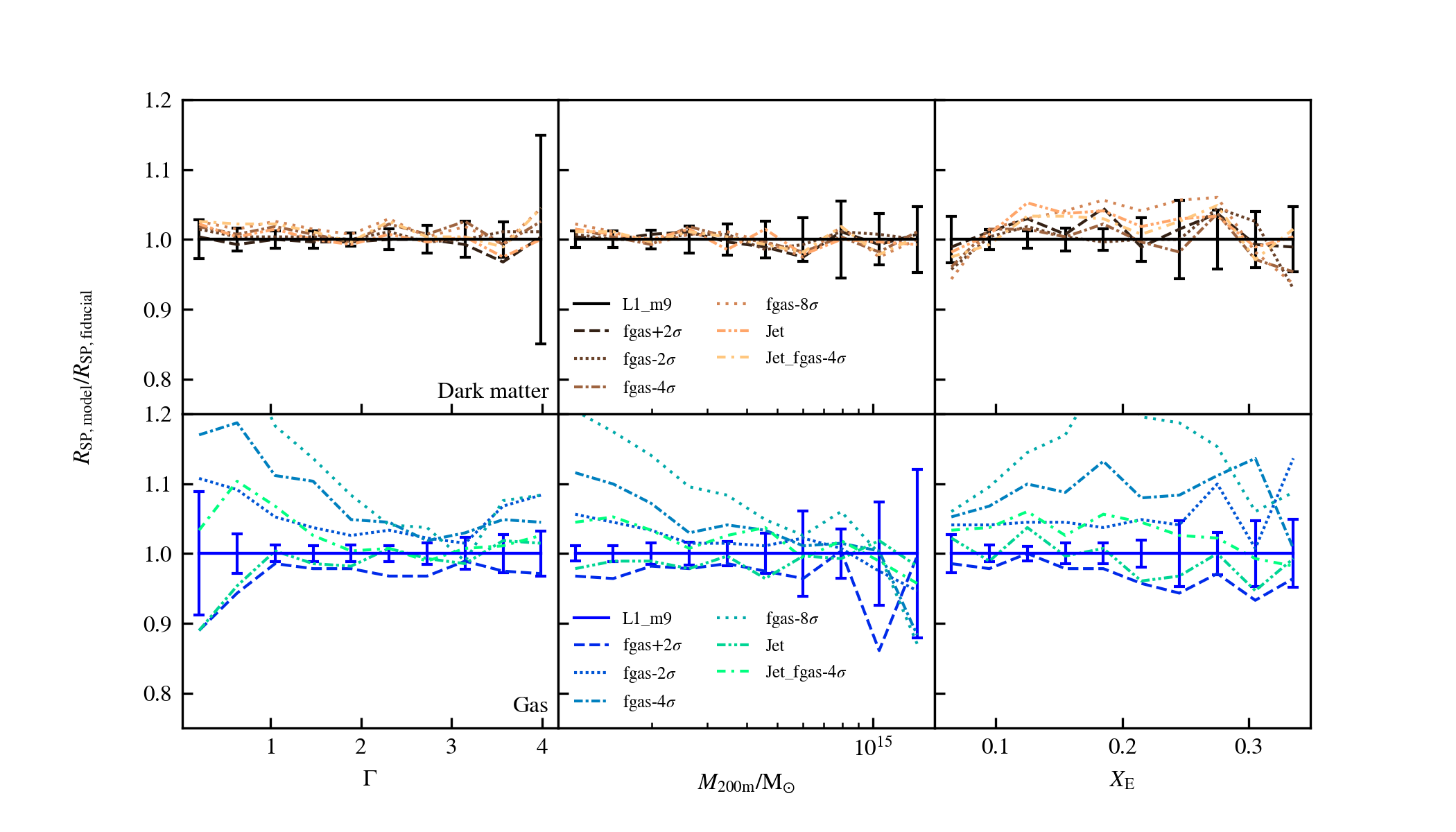}
    \caption{The variation between the splashback radius obtained for different stacking bins (from left to right, halos are binned based on their accretion rate, mass and energy ratio respectively) for different baryonic physics runs in comparison to the fiducial run, L1\_m9. The top row shows the splashback radius obtained from the dark matter density gradient profiles and the bottom row shows the radius of the feature reflected in the gas. The error bars show the uncertainty obtained by bootstrap resampling the clusters from the fiducial model.}
    \label{fig:baryonic_Rsp}
\end{figure*}

\subsubsection{Alternative baryonic models}

We now look at  the splashback feature obtained from dark matter and gas profiles from simulations that vary the resulting cluster gas fraction (models fgas+2, -2, -4 and -8$\sigma$) and feedback mechanism (Jet and Jet\_fgas-4$\sigma$, the latter also altering the cluster gas fraction). This allows us to probe whether the baryonic physics model used in a simulation affects the splashback feature and how sensitive it is to varying amounts of AGN feedback.

We compare the radius of the splashback minimum feature with different hydro runs relative to the fiducial run, L1\_m9, in Fig. \ref{fig:baryonic_Rsp}, with error bars showing an estimate on the amount of noise we expect due to sampling calculated by bootstrapping the clusters. The top row shows the differences between the splashback feature found in the dark matter density profiles and the second row shows the same for minima in the gas. We find, in agreement with \citet{ONeil2021TheSimulations}, that the location of the splashback radius itself (i.e. in the dark matter density) is mostly unaffected by the baryonic physics model used. Even in the most extreme cases shown, the splashback radius only varies by about 5-6 per cent from the fiducial run but the difference is not significantly larger than the sampling uncertainty.

However, we find that the baryonic physics model used has a much larger effect on the radius of the gas minimum density gradient. The stronger AGN feedback runs, e.g. fgas-4$\sigma$ and fgas-8$\sigma$, tend to have minima at larger radii and similarly the weaker AGN run, fgas+2$\sigma$, at smaller radii. The "Jet" runs result in the gas minima occurring at smaller radii. The location of the minimum is strongly dependent on the amount of feedback, i.e. energy given off by the cluster AGN, within a cluster and increased feedback effectively "blows" out the minimum to a larger radius. Lower mass clusters are more susceptible to the effects of feedback so this is most obvious in the middle panel where we can see models with higher levels of feedback have a stronger effect on the radius of the minima of lower mass clusters. This shows that the location of the minimum in the gas is not strictly defined by the location of the splashback (which is set by gravitational physics) and various other hydrodynamical effects within the gas can easily shape and move it away from the splashback radius.

%Fig. \ref{fig:baryonic_gamma} is similar to Fig. \ref{fig:baryonic_Rsp} but instead of comparing the radii of the minima, we compare the minimum gradient values. Again, we find that the dark matter results (top) are largely unaffected by variations in the baryonic physics models. However, the models with increased feedback have shallower minima in the gas profiles. The same process that moves the gas minima to higher radii also broadens the region containing the sharpest change in the gas density gradient and also how fast the density decreases.

\section{Results from projected profiles}
\label{sec:projection}

Identifying the splashback radius in 3D profiles is useful to check how it is affected by cluster properties, baryonic physics and statistical effects. However, observers will only obtain projected images of clusters and so the resulting measurement of the splashback radius will be different. We discuss the effects of projection on the log-slope density profiles as well as pseudo-observable profiles to investigate the differences between the splashback radii obtained from 3D density profiles and the radii of the minima of observable profiles.

\subsection{Projected splashback radius}
\label{sec:projection:direct}
%Discuss observation paper who have done this to obtain 3D radius
%Discuss simulation papers and their results from projection 

We can calculate the radius where we expect to find splashback features in projected profiles by fitting a model to the stacked 3D dark matter density profile and projecting that model. Following \citet{Diemer2014DEPENDENCERATE}, we fit the following model for the dark matter density:
\begin{equation}
    \rho(r) = \rho_{\rm{inner}} \times f_{\rm{trans}} + \rho_{\rm{outer}},
\end{equation}
where
\begin{equation}
    \rho_{\rm{inner}} = \rho_{\rm{s}} \exp \left( -\frac{2}{\alpha} \left[ \left( \frac{r}{r_{\rm{s}}} \right) ^{\alpha} - 1 \right] \right)
\end{equation}
is the Einasto model describing the inner halo,
\begin{equation}
    f_{\rm{trans}} = \left[ 1 + \left( \frac{r}{r_{\rm{t}}} \right) ^{\beta} \right]^{- \frac{\gamma}{\beta}}
\end{equation}
models the transition region and
\begin{equation}
    \rho_{\rm{outer}} = \rho_{\rm{m}} \left[ b_{e} \left( \frac{r}{5 R_{\rm{200m}}} \right) ^{-S_{e}} + 1 \right],
\end{equation}
models the outer density profile.
Once the ideal free parameters, $\{\rho_{\rm{s}}, r_{\rm{s}}, r_{\rm{t}}, \alpha, \beta, \gamma, b_{e}, S_{e} \}$, have been fitted to each dark matter density profile, we project the model using,
\begin{equation}
    \Sigma (R) = 2 \int_{R}^{5 R_{\rm{200m}}} \frac{\rho(r) r \rm{d} r}{\sqrt{r^2 - R^2}},
\end{equation}
and identify the minimum from the gradient of the resulting projected profile. We then use this as the expected splashback radius in 2D to compare with observable profiles. Observational works such as \citet{More2016DetectionClusters} and \citet{Zurcher2019Clusters} have shown that the splashback radius found from projected profiles is smaller than the radius found in 3D. In this work, we find that the minima in these projected gradient profiles are located on average at roughly 0.82$\pm$0.03 times that of the 3D splashback radii.

\begin{figure*}
    \centering
    \includegraphics{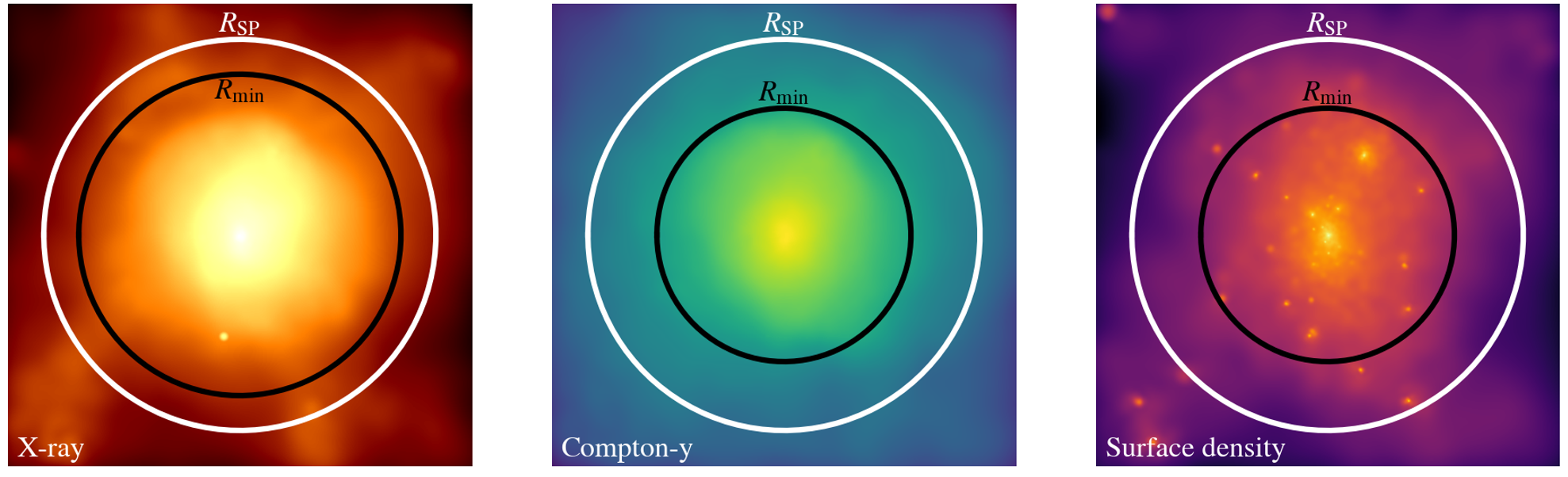}
    \caption{Maps of an example FLAMINGO cluster with mass $M_{\rm{200m}} = 1.97 \times 10^{14}\,  \rm{M_{\odot}}$ and accretion rate $\Gamma=1.99$ at $z=0$. From left to right, these maps show the emission measure, Compton-$y$ and total mass surface density of the cluster. Each map shows the location of the splashback radius ($R_{\rm{SP}} = 1.27 R_{\rm{200m}})$ obtained from the cluster's 3D dark matter density gradient profile in white and the location of the minimum in the gradient profile of the respective 2D observable in black. We find the location of the minima to be $R_{\rm{min}}=1.04, 0.82$ and $0.94 R_{\rm{200m}}$ for the emission measure, Compton-y and total mass surface density respectively.}
    \label{fig:maps}
\end{figure*}

\begin{figure*}
    \centering
    \includegraphics[trim = {0.8cm 0.5cm 1.5cm 1.5cm}, clip]{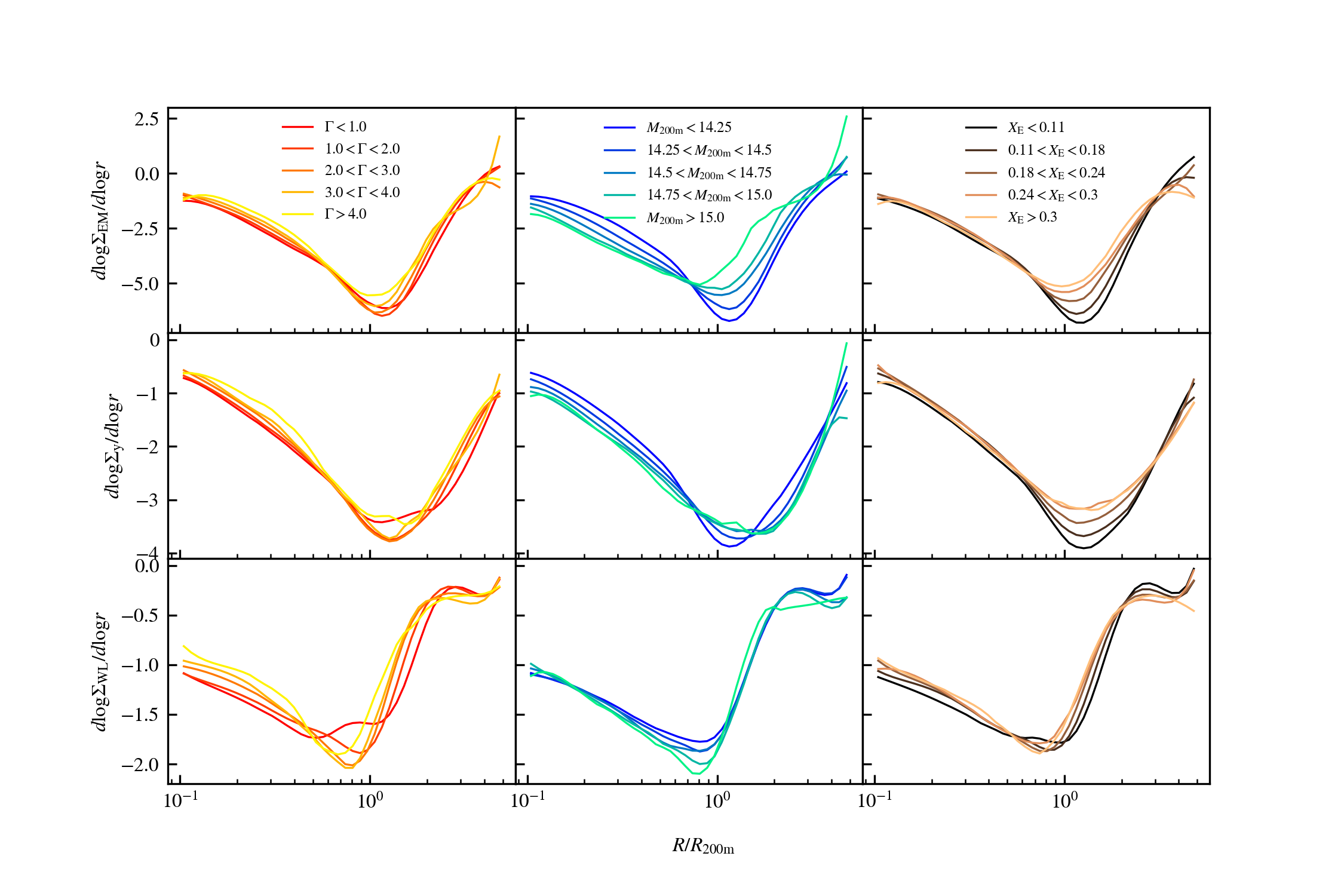}
    \caption{Stacked projected gradient profiles for the three observables tested in this work. Top: X-ray emission measure, middle: Compton-$y$, bottom: total mass surface density. These profiles have been stacked according to the mass accretion rate (Equation \ref{eq:accretion}, left column), mass (central column) and gas energy ratio (Equation \ref{eq:energy}, right column) of the clusters.}
    \label{fig:profs_2D}
\end{figure*}

\begin{figure*}
    \centering
    \includegraphics{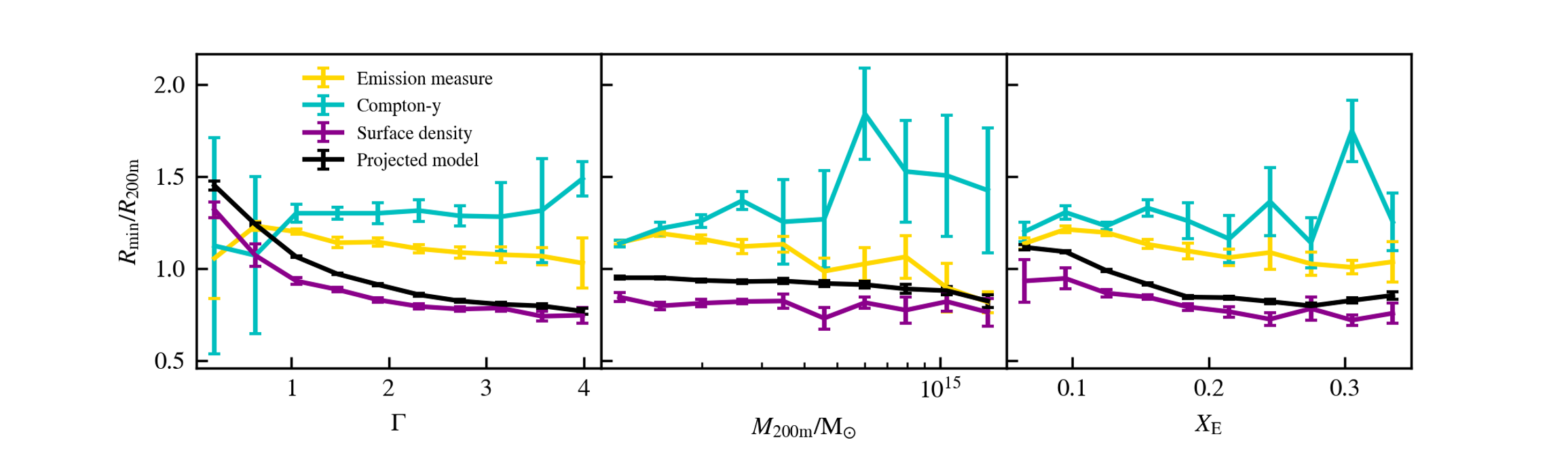}
    \caption{Comparison of the location of the minima of the three different observable gradient profiles used in this work (emission measure, gold; Compton-$y$, blue; and surface density, purple) as well as the expected 2D splashback feature obtained from projecting the 3D dark matter density profiles. Each panel compares different bins used to stack the profiles (left: accretion rate, centre: mass and right:energy ratio). Error bars show the 1 $\sigma$ uncertainty from bootstrap resampling.}
    \label{fig:params_2D}
\end{figure*}

\subsection{Observable profiles}
\label{sec:projection:obs}

As described in Section \ref{sec:sims:profiles}, profiles have been extracted from the FLAMINGO halos to broadly represent what can be obtained from observations. Each halo was projected three times, once for each perpendicular axis of the simulation, and azimuthally averaged emission measure (X-ray), Compton-$y$ (SZ) and total surface density (weak lensing) profiles were extracted from each projection with a total thickness of 10 $R_{\rm{200m}}$. In this section, we show how these profiles are affected by different stacking methods as well as if these observable profiles could be used to accurately measure the splashback radius. 

We show maps of an example FLAMINGO cluster of mass $M_{\rm{200m}} = 1.97 \times 10^{14} \rm{M_{\odot}}$ and accretion rate $\Gamma = 1.99$ in each of these three observables in Fig. \ref{fig:maps}. The maps include circles denoting the location of the 3D splashback radius for this cluster (1.27 $R_{\rm{200m}}$, in white) calculated from the dark matter density gradient profiles and the location of the minimum, $R_{\rm{min}}$, in each observable's gradient profile (shown in black). The 3D splashback radius, as well as $R_{\rm{min}}$, in both emission measure and in surface density are approximately located where one would expect for a cluster of this mass (see Figs. \ref{fig:params} and \ref{fig:params_2D}). However, $R_{\rm{min}}$ for the Compton-$y$ profile is smaller than we find in the stacked profiles. This is the result of looking at a singular cluster rather than a stacked cluster. Compton-$y$ profiles of individual clusters often contain multiple minima and so the radius of the deepest minimum is found at smaller radii than in stacked profiles where it is smoothed by the larger shock radius feature.

The observable gradient profiles stacked according to the accretion rate, mass and energy ratio are shown in Fig. \ref{fig:profs_2D}. The log-slope of the emission measure profiles show a much deeper minimum than the log-slope of the gas density (Fig. \ref{fig:params}), because of the density-squared dependence of the emission measure. The SZ profiles are similar to what we obtained from the 3D pressure profiles (Fig. \ref{fig:profile_comparison}), however we find much broader minima in the SZ due to projection effects. As the mass of the cluster is dominated by dark matter, we expect that the surface density gradient profiles are similar to the gradient profiles for the 3D dark matter density \citep{Xhakaj2020HowRate}. We find that this is true, and that the location of the minima moves to smaller radii and is broader than in the 3D dark matter profiles.

\begin{figure*}
    \centering
    \includegraphics[trim = {1cm 0.2cm 1.8cm 1cm}, clip]{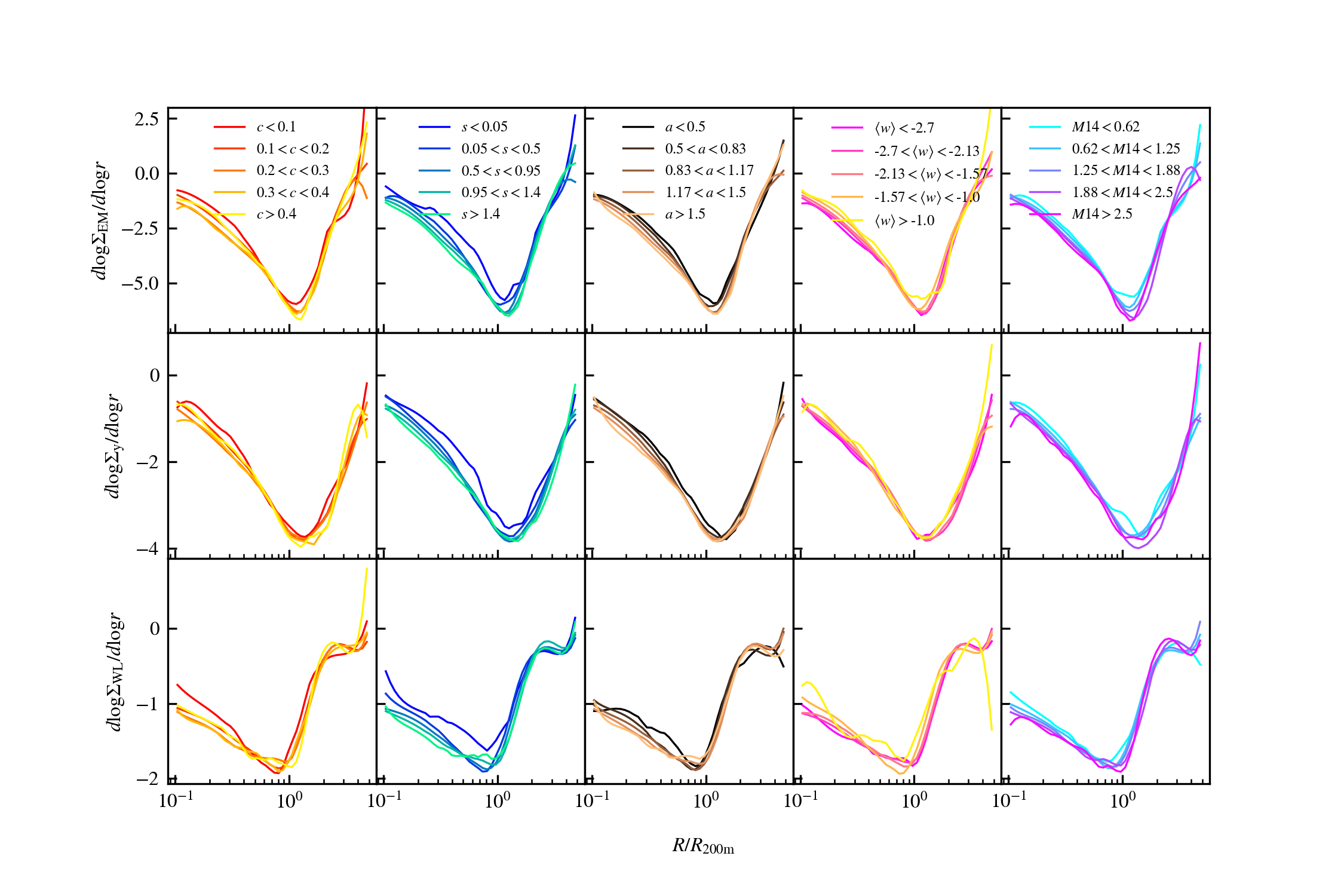}
    \caption{Stacked projected gradient profiles of the three observables tested in this work, top: X-ray emission measure, middle: Compton-$y$, bottom: total mass surface density obtained from weak lensing. These profiles have been stacked according to the four morphology criteria and the magnitude gap, see Section \ref{sec:stacking:obs}. From left to right, the plot shows bins in concentration, symmetry, alignment, centroid shift and magnitude gap. The clusters shown are restricted to the range $10^{14.2} < M_{\rm{200m}} < 10^{14.4}\,  \rm{M_{\odot}}$.}
    \label{fig:morph_all}
\end{figure*}

\begin{figure*}
    \centering
    \includegraphics{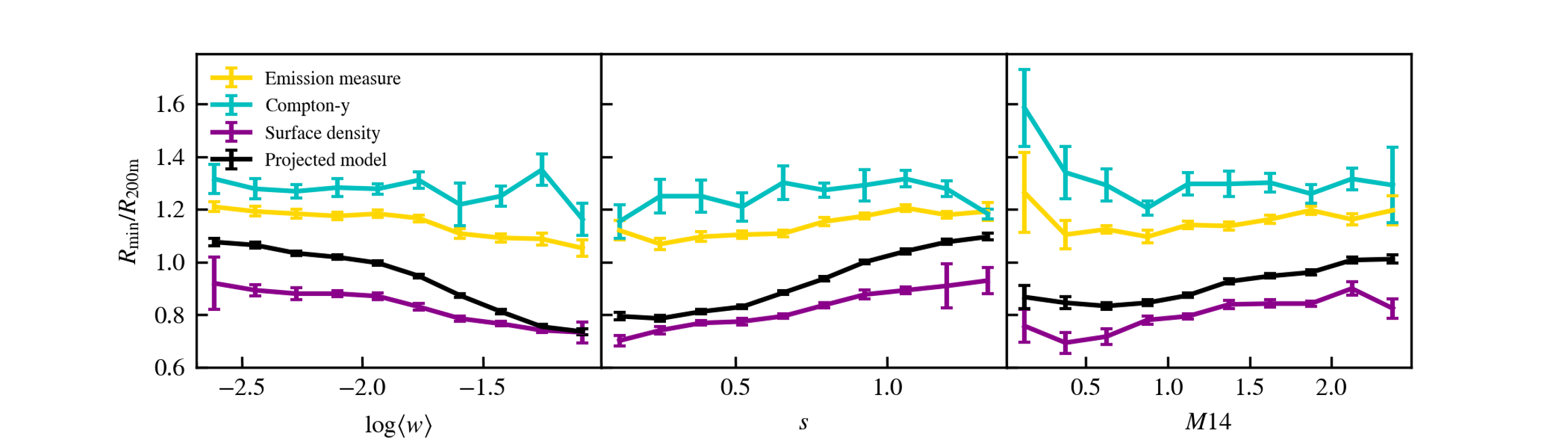}
    \caption{Same as Fig. \ref{fig:params_2D} but instead compares the parameter dependence of the radius of the minimum in the gradient profile with the centroid shift (Equation \ref{eq:centroid}, left), symmetry statistic (Equation \ref{eq:symmetry}, middle) and magnitude gap (right).}
    \label{fig:obs_dependence}
\end{figure*}

We study the radius of the minimum gradient more closely in Fig. \ref{fig:params_2D}. We also include the location of the minima we expect from projecting the same sample of 3D dark matter density profiles (black line). Overall, there is little correspondence between the radius of the minima in the three observables. In general, the surface density matches that of the projected model, again as expected as the surface density will be dominated by dark matter. We consistently identify minima in the emission measure and Compton-y profiles but these occur at a larger radius than the splashback radius. The differences between the locations of these minima and the splashback radius may have implications for the underlying physics of the components each of the observables probe.

Furthermore, in Fig. \ref{fig:params_2D}, we find that there is a mild anti-correlation between the accretion rate and the radius of the minimum in the total surface density gradient, similar to what we found for the dark matter density in Fig. \ref{fig:params} but with a slightly less steep curve at low accretion rates. We find that the minimum radius is smaller in the surface density than for the projected 3D dark matter density due to the gas contribution. The minimum of the gas density gradient tends to occur at smaller radii than the dark matter density (see Fig. \ref{fig:params}) and so when these are combined in the total surface density, the radius of the gradient minimum is reduced with respect to the dark matter alone. 

Fig. \ref{fig:params_2D} also shows that there is no strong correlation between $R_{\rm{min}}$ for the surface density and the mass or the energy ratio. In addition, we find that the radius of the minimum in the emission measure depends on the mass of the cluster. High-mass clusters have minima at smaller radii, at similar locations as the dark matter splashback radii. Fig. \ref{fig:params} showed that the minima found in 3D gas density gradient profiles do not show much of a correlation with either the accretion rate or the energy ratio. Similarly, we find that $R_{\rm{min}}$ for the emission measure has a weak correlation with both quantities. Emission measure is expected to scale roughly with density squared and so we expect similar trends between the emission measure and 3D gas density. In addition, while there is not much of a trend between the minima in the Compton-$y$ gradient and the accretion rate or energy ratio, the radius of the minimum generally increases with mass. 

This is at odds with the minimum in the pressure profiles (Fig. \ref{fig:profile_comparison}) which decreases slightly for increasing mass, though this effect is not strong over the mass range used. In Fig. \ref{fig:profs_2D}, we saw that the Compton-$y$ gradient profiles of the high mass clusters have a broader minimum than at lower masses. In the pressure, it is common to have additional minima in the gradient profiles of individual higher mass clusters, but as these profiles have been stacked, the extra minima have been smoothed out to create a single broader minimum. In the projected Compton-$y$ gradient profiles, this broadening of the minima tends to lead to $R_{\rm{min}}$ being identified at larger radii but with an increased uncertainty. However, in the stacked 3D pressure profiles, Fig. \ref{fig:profile_comparison}, the minima at the larger radii are not as significant. The radii of minima are reduced by projection and so the outer minima are further out in 3D and have less of an effect on the pressure profiles around the splashback radius.

\subsubsection{Stacking profiles using observables}

While the splashback radius has a strong dependence on the accretion rate of the host halo, the accretion rate is not an observable property of clusters. Instead, we aim to find an observable proxy that would allow the cluster profiles to be stacked appropriately. In Fig. \ref{fig:morph_all}, we show the three projected gradient profiles (top: emission measure, middle: Compton-$y$, bottom: surface density) from clusters in the mass range $10^{14.2} < M_{\rm{200m}}/\rm{M_{\odot}} < 10^{14.4}$. These were stacked in bins of the five properties introduced in Section \ref{sec:stacking:obs} (left to right: concentration, symmetry, alignment, centroid shift and magnitude gap). When stacking according to the surface brightness concentration, we find that the profiles are all very similar, particularly towards the minimum. The differences appear in the core of the cluster, which is to be expected as the concentration parameter probes differences in the centre rather than the outskirts of the cluster. The profiles separate more when stacked according to the symmetry statistic, so much so that a small second caustic feature appears in the surface density gradient profiles of the most regularly shaped clusters ($s>1.4$).
Due to the strong correlation between the symmetry statistic and the accretion rate (see Fig. \ref{fig:parameter_corner}), we expect there to be a large number of low accretion rate clusters within the bins with high $s$, increasing the likelihood for the second caustic to appear. However, this feature is not as clearly defined as seen previously for the accretion rate because the symmetry statistic and accretion rate do not correlate perfectly. 

Motivated by the stronger correlations between the accretion rate and the symmetry statistic, centroid shift and magnitude gap, we plot in Fig. \ref{fig:obs_dependence} how the extracted minimum radii of the three projected profiles varies with each of these statistics. A smaller symmetry and a lower centroid shift both correspond to more dynamically disturbed clusters and we find that, for both statistics, the radius of the minimum extracted from the surface density and emission measure gradient profiles decreases for more dynamically disturbed clusters, matching the more dynamically disturbed clusters with higher accretion rates and smaller splashback radii.

\citet{Shin2022WhatProperties} proposed that the accretion rate of halos negatively correlates with the magnitude gap between the BCG and the brightest satellite galaxy. Over time, satellites get stripped of matter and tidally disrupted, and larger galaxies sink to the centres of clusters due to dynamical friction. It has been shown that this effect is most prominent for the largest satellites, meaning that over time the brightness gap increases as a halo grows. We measured the magnitude gap following \citet{Farahi2020AgingHaloes}, see Section \ref{sec:stacking:obs}. We find a mild, negative correlation between the accretion rate and the magnitude gap (this was found to be slightly dependent on the resolution of the simulation, see Appendix \ref{app:mag_gap}) and therefore expect a positive correlation between the magnitude gap and the splashback radius. The rightmost column of Fig. \ref{fig:obs_dependence} compares $R_{\rm{min}}$ obtained from the observable profiles with the magnitude gap. We find that there is a slight correlation between the two, most prominent in the surface density profiles. However, we find almost no correlation between the magnitude gap and the minima found in either the emission measure or Compton-$y$ gradient profiles. %We expect that the surface density profiles are able to recover the splashback feature most accurately as it is dominated by dark matter density.

\section{Conclusions}
\label{sec:conclusions}

In this work we have used clusters from the FLAMINGO cosmological hydrodynamical simulations to obtain dark matter density, gas density, gas pressure and gas entropy profiles. From these we have extracted the location of the minimum local gradient to identify the splashback radius from the dark matter or a potentially matching feature in the gas profiles. We investigated the effect of stacking the profiles in a variety of ways and how the location of the radius changes when using projected profiles. Our results can be summarised as follows:

\begin{itemize}
    \item The splashback radius identified using the dark matter density gradient profile has a strong anti-correlation with the accretion rate, in agreement with previous works (see Figs. \ref{fig:profile_comparison} and \ref{fig:params}). 
    \item The splashback radius has a weak negative mass dependence. Previous works have suggested that this is due to the connection between cluster mass and accretion rate \citep{Diemer2014DEPENDENCERATE}. We find a weak correlation (see Fig. \ref{fig:parameter_corner}) and when stacking in bins of both mass and accretion rate (Fig. \ref{fig:grid_mass_acc}), while the accretion rate has obvious qualitative effects on the location of the splashback radius, there is not a significant mass dependence.
    \item We also identify a minimum in the cluster gas density gradient profiles, approximately corresponding to the dark matter splashback radius, but the radius of this minimum has a stronger mass dependence (Figs. \ref{fig:profile_comparison} and \ref{fig:grid_mass_acc}). In addition, we find a similar feature in the gas pressure gradient, with no further minima indicating potential shock features at higher radii. However, the gas entropy gradient profiles have minima in the region we expect to find a shock feature in the gas, i.e. 2-3 $R_{\rm{200m}}$ \citep{Aung2021ShockClusters}.
    \item Comparison between hydrodynamical and dark matter only simulations finds minimal difference in the recovered splashback radii (see Fig. \ref{fig:dmo_scatter}).
    \item Altering the astrophysical parameters used in the simulation results in an essentially unchanged splashback radius. However the radius of the gas minima is much more sensitive to the astrophysics models within the simulation, reducing the correspondence between the splashback and gas minimum (Fig. \ref{fig:baryonic_Rsp}).
    \item In Section \ref{sec:projection}, we investigate the effects of projection on the recovered splashback radius. Fitting the dark matter density following \citet{Diemer2014DEPENDENCERATE} to 3D profiles out to 5 $R_{\rm{200m}}$ and projecting the fitted profiles results in the minima occurring at 0.8 times that of the minima in the 3D profiles (Section \ref{sec:projection:direct}).
    \item We find that the minimum of the total mass surface density gradient profile has a similar radius to what we expect from the projected 3D dark matter profiles. However, although similar, it is found at a systematically smaller radius due to the contribution of the gas (see Fig. \ref{fig:params_2D}).
    \item We find that the minima of the observable profiles (emission measure, Compton-$y$ and total surface density) are dependent on the morphology measures used to stack the profiles: to varying degrees more regular clusters tend to have larger splashback radii (see Fig. \ref{fig:obs_dependence}). In addition, the radius of the surface density minima most closely resembles where we expect the splashback radius to be after projection. This is to be expected, as the surface density is dominated by dark matter. However, the minima of the emission measure and Compton-$y$ gradient profiles are located at substantially larger radii and are nearly independent of the dynamical state.
\end{itemize}

The gas density and pressure gradient profiles of the FLAMINGO clusters demonstrate that there exists a minimum at approximately the splashback radius. There are similarly also minima at similar radii in the gradient profiles of gas observables such as the emission measure and Compton-$y$. However, due to the differing physics governing the gas and dark matter motions, it is unclear whether these minima are a true reflection of the splashback radius or coincidentally located at a similar radius. Observational works such as \citet{Anbajagane2023CosmologicalACT} have started to identify minima at radii similar to what we have found in Compton-$y$ gradient profiles of clusters. Future optical/IR, X-ray and SZ observations will be useful to compare with simulations and further our understanding of the behaviour of baryons and dark matter in cluster outskirts.

\section*{Acknowledgements}

This work used the DiRAC@Durham facility managed by the Institute for Computational Cosmology on behalf of the STFC DiRAC HPC Facility (www.dirac.ac.uk). The equipment was funded by BEIS capital funding via STFC capital grants ST/K00042X/1, ST/P002293/1, ST/R002371/1 and ST/S002502/1, Durham University and STFC operations grant ST/R000832/1. DiRAC is part of the National e-Infrastructure.

We also wish to thank the Science and Technologies Facilities Council for providing studentship support for IT. This work is partly funded by research programme Athena 184.034.002 from the Dutch Research Council (NWO).

%%%%%%%%%%%%%%%%%%%%%%%%%%%%%%%%%%%%%%%%%%%%%%%%%%

\section*{Data Availability}
The data shown in the plots within this article are available upon reasonable request to the corresponding author. The code used in the analysis is publicly available on the corresponding author’s GitHub repository (\href{https://github.com/imogentow/splashback}{https://github.com/imogentow/splashback}). The data from the FLAMINGO suite will be made public once practically feasible given the challenging size of the data sets. In the mean time, the FLAMINGO data can be accessed upon requests to the authors.


\bibliographystyle{mnras}
\bibliography{references, references2}

%%%%%%%%%%%%%%%%%%%%%%%%%%%%%%%%%%%%%%%%%%%%%%%%%%

%%%%%%%%%%%%%%%%% APPENDICES %%%%%%%%%%%%%%%%%%%%%

\appendix

\section{Effects of resolution on magnitude gap}
\label{app:mag_gap}

The magnitude gap between the BCG and the fourth brightest galaxy within a cluster correlates with the cluster's accretion rate and therefore, its splashback radius. However, galaxies within the cosmological simulation are poorly resolved. In the case of smaller halos or halos with much more mass concentrated in the main halo, the fourth brightest galaxy can be difficult to resolve and therefore our estimates for the magnitude difference between the two galaxies may be more uncertain. The FLAMINGO suite contains multiple resolutions of the same simulation box, L1\_m9 for the standard resolution used in this work (gas particle mass $m_{\rm{gas}} \approx 10^9\,  \rm{M_{\odot}}$ and L1\_m8 a higher resolution where the particles are a factor of eight less massive (note the models are calibrated separately but to the same observable). This allows us to directly test whether the resolution of the simulation has an effect on the retrieved magnitude gap and its correlation with the accretion rate or splashback radius.

\begin{figure}
    \centering
    \includegraphics{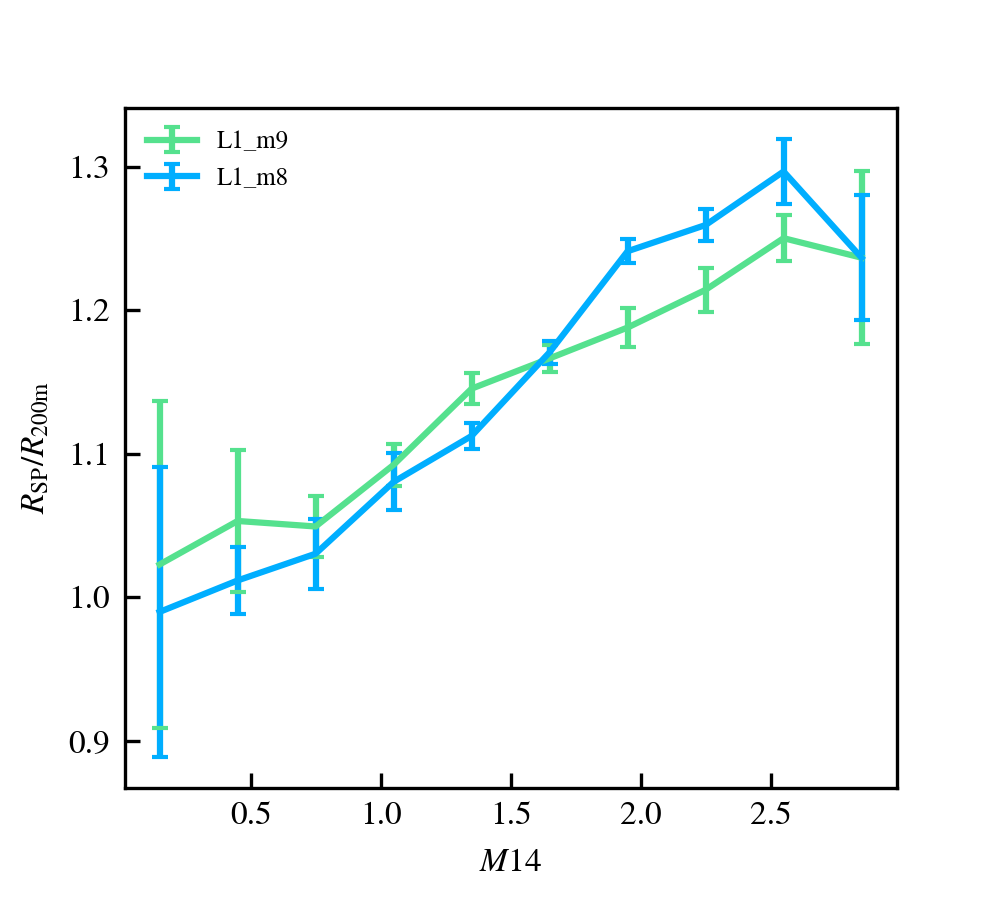}
    \caption{Comparison between the splashback obtained from dark matter density profiles when stacked according to the magnitude gap for two different resolution simulations. The clusters all have a mass of $M_{\rm{200m}} > 10^{14}\,  \rm{M_{\odot}}$. The error bars show the expected 1 sigma errors due to resampling.}
    \label{fig:res_comparison}
\end{figure}

We find that the Pearson correlation coefficient between the accretion rate and the magnitude gap increases in the higher resolution simulation, from -0.35 in L1\_m9 to -0.48 in L1\_m8. We plot the splashback radius obtained from the dark matter density gradient profile stacked according to the magnitude gap in Fig. \ref{fig:res_comparison}. We find that there is a slightly steeper relationship between the magnitude gap and the splashback radius in the higher resolution simulation, but the two results are qualitatively similar.

\begin{figure*}
    \centering
    \includegraphics{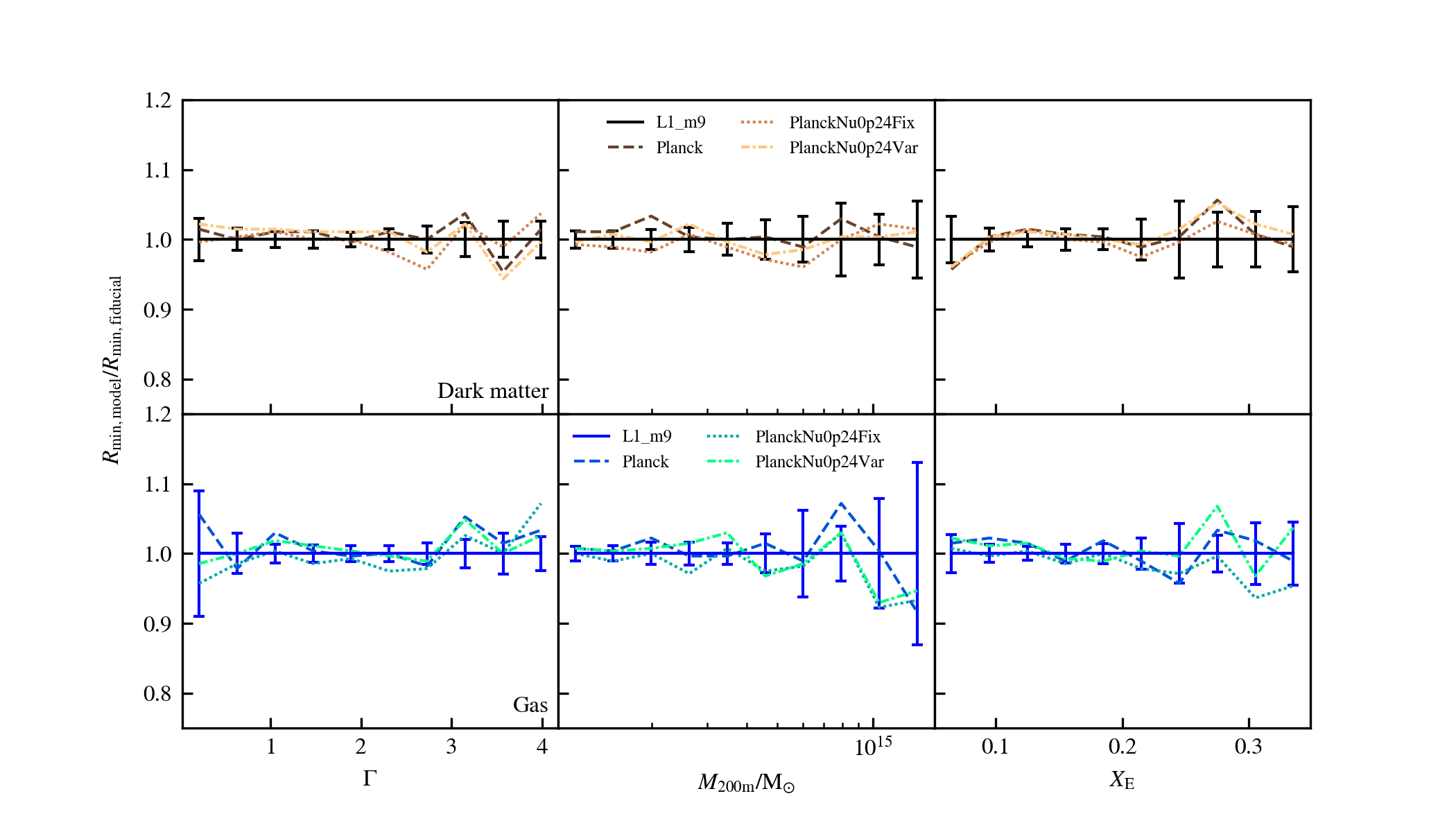}
    \caption{The variation between the splashback radius obtained for different stacking bins (from left to right, halos are binned based on their accretion rate, mass and energy ratio) for runs with different cosmological models in comparison to the fiducial run, L1\_m9. The top row shows the splashback radius obtained from the dark matter density gradient profiles and the bottom row shows the radius of the feature in the gas.}
    \label{fig:cosmo_Rsp}
\end{figure*}

%\begin{figure*}
%    \centering
%    \includegraphics{Figures/parameter_dependence_cosmo_gamma.png}
%    \caption{Same as Fig. \ref{fig:cosmo_Rsp} but instead looks at the variation of the depth of the splashback feature, $\gamma$, rather than its radius.}
%    \label{fig:cosmo_gamma}
%\end{figure*}

\section{Alternative cosmological models}
\label{app:cosmo}

In addition to the varying baryonic physics models in Section \ref{sec:models}, we also investigate the effects of varying the cosmological model on the splashback radius. Fig. \ref{fig:cosmo_Rsp} shows the variation of the minima in the dark matter (top panels) and gas (bottom panels)  density gradient profiles relative to the fiducial model when using different cosmological models. We find that the splashback radius extracted from the dark matter density profiles is essentially unaffected by the change in cosmological model of the simulation, agreeing with the cosmological independence found in \citet{Diemer2017TheCosmology}. In addition, we find that the location of the gas minimum is less affected by the change in cosmological model as expected as the cosmological model has less of an effect on the gas physics. Additionally, for both the gas and the dark matter, the effect of the variation of the cosmological model on the depth of the minima is only of the order of a few percent. Thus the depth of the minimum is also much more sensitive to the baryonic physics than to the cosmological model of the simulation.

%%%%%%%%%%%%%%%%%%%%%%%%%%%%%%%%%%%%%%%%%%%%%%%%%%

% Don't change these lines
\bsp	% typesetting comment
\label{lastpage}
\end{document}